\documentclass[preprint]{vgtc}               % preprint
\graphicspath{{figures/}{pictures/}{images/}{./}} % where to search for the images

\usepackage{times}                     % we use Times as the main font
\usepackage{tabu}                      % only used for the table example
\usepackage{booktabs}                  % only used for the table example
\usepackage{lipsum}                    % used to generate placeholder text
\usepackage{mwe}                       % used to generate placeholder figures
\usepackage{subfigure}
\usepackage{graphicx}
\usepackage{subcaption} % for subfigures
\usepackage{caption} % for customizing captions
\usepackage{amsmath}
\usepackage{amssymb}
\usepackage{algorithm}
\usepackage{algpseudocode}
\DeclareUnicodeCharacter{03BB}{$\lambda$}

\newcommand{\changed}[1]{{\textcolor{black}{#1}}}

\usepackage{mathptmx}                  % use matching math font

\onlineid{1188}

\vgtccategory{Research}

\vgtcinsertpkg

\ieeedoi{10.1109/VIS55277.2024.00070}

\title{Topological Separation of Vortices}

\author{Adeel Zafar\thanks{e-mail:azafar3@uh.edu} %
\and Zahra Poorshayegh\thanks{e-mail:zpoorsha@uh.edu} %
\and Di Yang\thanks{e-mail:diyang@uh.edu} %
\and Guoning Chen\thanks{e-mail:gchen16@uh.edu}}
\affiliation{\scriptsize University of Houston}

\abstract{
Vortices and their analysis play a critical role in the understanding of complex phenomena in turbulent flows.
Traditional vortex extraction methods, notably region-based techniques, often overlook the entanglement phenomenon, resulting in the inclusion of multiple vortices within a single extracted region. Their separation is necessary for quantifying different types of vortices and their statistics. In this study, we propose a novel vortex separation method that extends the conventional contour tree-based segmentation approach with an additional step termed ``layering''. Upon extracting a vortical region using specified vortex criteria (e.g., $\lambda_2$), we initially establish topological segmentation based on the contour tree, followed by the layering process to allocate appropriate segmentation IDs to unsegmented cells, thus separating individual vortices within the region. However, these regions may still suffer from inaccurate splits, which we address statistically by leveraging the continuity of vorticity lines across the split boundaries. Our findings demonstrate a significant improvement in both the separation of vortices and the mitigation of inaccurate splits compared to prior methods.
} % end of abstract


\begin{document}
\setlength{\baselineskip}{0.99 \baselineskip}
\setlength{\abovedisplayskip}{1pt}
\setlength{\belowdisplayskip}{1pt}
\setlength{\abovedisplayshortskip}{2pt}
\setlength{\belowdisplayshortskip}{-2pt}
\setlength{\belowcaptionskip}{2pt}
\setlength{\abovecaptionskip}{2pt}
\setlength{\textfloatsep}{5pt}
\setlength{\floatsep}{2pt}
\setlength{\intextsep}{2pt}


\firstsection{Introduction}

\maketitle

%% \section{Introduction} %for journal use above \firstsection{..} instead
% Introduction comes here
% Vortices play a crucial role in many fluid phenomena, spanning applications from aerodynamics to environmental fluid dynamics. These dynamic structures arise from the interaction of fluid particles, forming swirling patterns that influence heat and mass transfer, mixing, and flow stability. Vortices vary in size, shape, and intensity, ranging from small-scale eddies to large, coherent structures that dominate flow patterns. Vortices can undergo complex interactions, including merging, splitting, and stretching, which further amplify their impact on flow dynamics. %Understanding the intricate dynamics of vortices is essential for predicting and controlling fluid behavior in various applications, from improving the efficiency of aircraft wings to optimizing the mixing processes in industrial reactors.
% In turbulent flows, vortices are particularly prevalent, contributing to the chaotic and unpredictable nature of the motion. 
\label{sec:intro}

\begin{figure}[t]
 \centering
    \begin{minipage}{0.5\linewidth}
    \centering
        \begin{subfigure}
             \centering
             \includegraphics[width=1\linewidth]{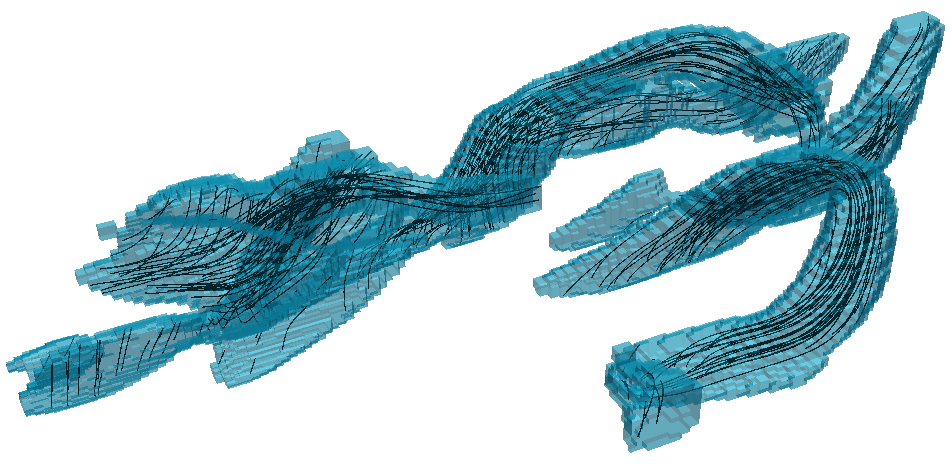}
             \vskip -15pt
             \caption*{(a)}
        \end{subfigure}
    \end{minipage}%
    \begin{minipage}{0.5\linewidth}
    \centering
        \begin{subfigure}
             \centering
             \includegraphics[width=1\linewidth]{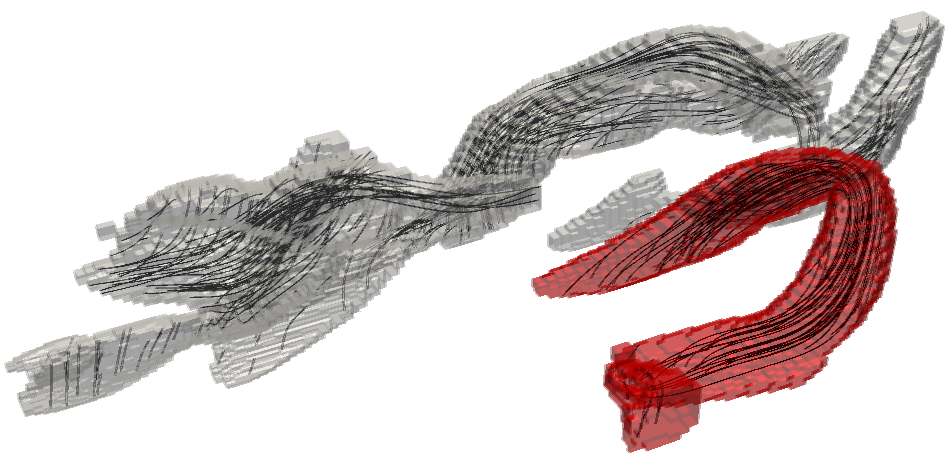}
             \vskip -15pt
             \caption*{(b)}
        \end{subfigure}
    \end{minipage}%
    \vskip 0.5pt
    \begin{minipage}{0.5\linewidth}
    \centering
        \begin{subfigure}
             \centering
             \includegraphics[width=1\linewidth]{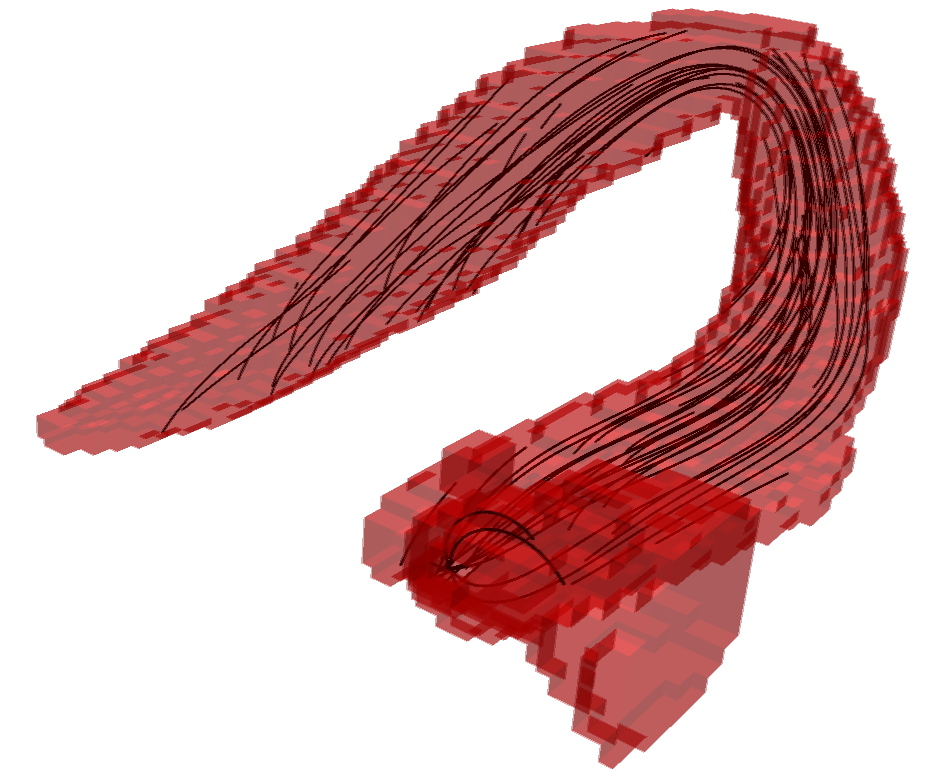}
             \vskip -15pt
             \caption*{(c)}
        \end{subfigure}
    \end{minipage}%
    \begin{minipage}{0.5\linewidth}
    \centering
        \begin{subfigure}
             \centering
             \includegraphics[width=1\linewidth]{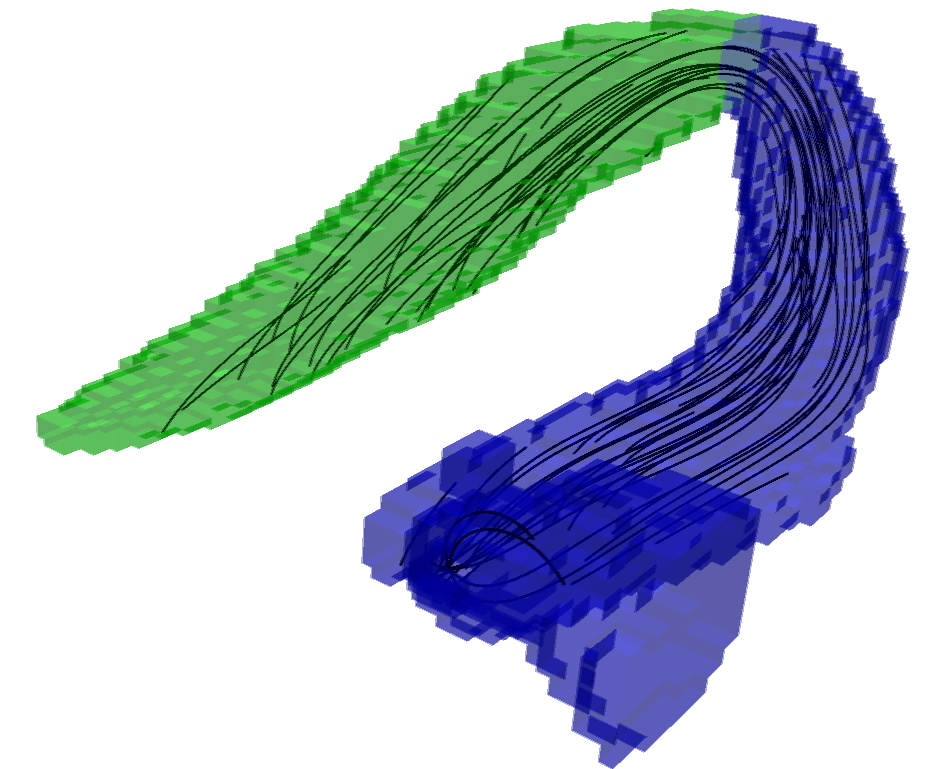}
             \vskip -15pt
             \caption*{(d)}
        \end{subfigure}
    \end{minipage}%
    \vskip -5pt
 \caption{(a) shows a vortical region (light-blue) extracted with the region growing~\cite{adeelhairpin2023} using $\lambda_2$ criterion. Variance in the patterns of vorticity lines (black) indicate the presence of multiple vortices. (b) shows a hairpin vortex (red) correctly getting separated from the rest of the vortices. (c) shows the zoomed in version of the hairpin vortex. (d) shows the inaccurate split of the hairpin vortex.}
 \label{fig_1_1}
\end{figure}

Region-based vortex extraction methods (e.g., $\lambda_2$~\cite{jeong1995identification}, $Q$~\cite{hunt1987vorticity}, $\lambda_{ci}$~\cite{ZHOU_ADRIAN_BALACHANDAR_KENDALL_1999}) are commonly employed to identify and analyze vortical structures within flow fields. However, in turbulent environments characterized by high Reynolds numbers, vortices often exhibit complex configurations and intricate interactions, leading to entanglement phenomena. This entanglement complicates the accurate extraction and characterization of vortices, as a single extracted region may contain multiple intertwined vortices as shown in \cref{fig_1_1}. Addressing this challenge is essential for gaining insights into specific vortex behaviors and their role in flow dynamics. In this work, we improve the vortex separation technique from~\cite{adeelhairpin2023} which presents several limitations.

\begin{figure}[t]
 \centering 
   \begin{minipage}{0.5\linewidth}
    \centering
        \begin{subfigure}
             \centering
             \includegraphics[width=1\linewidth]{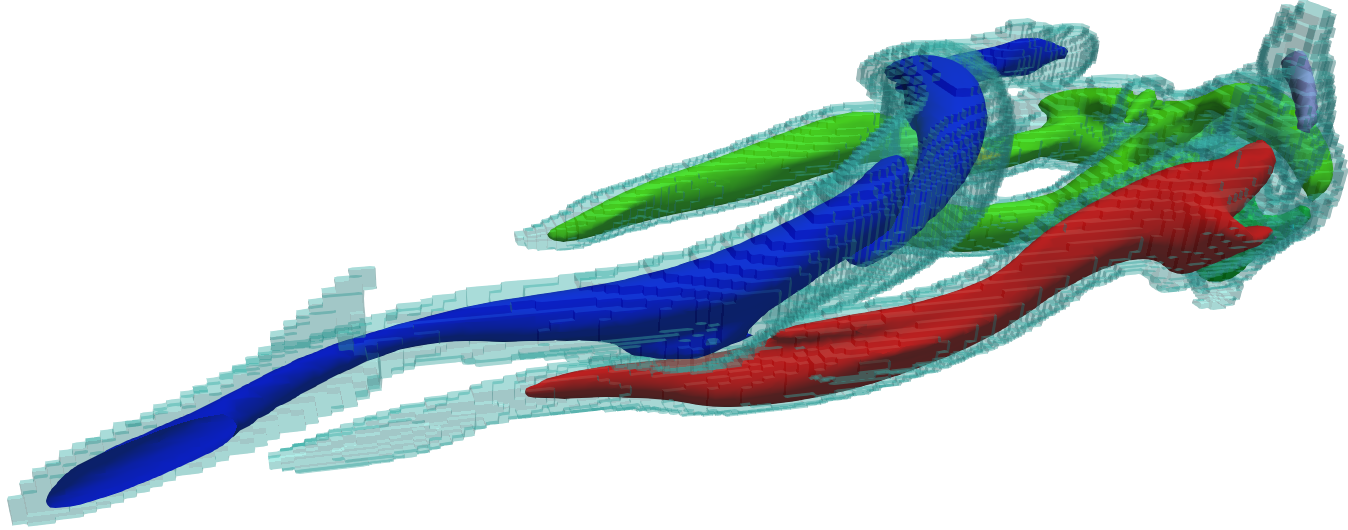}
             \vskip -10pt
             \caption*{(a)}
        \end{subfigure}
    \end{minipage}%
    \begin{minipage}{0.5\linewidth}
        \centering
        \begin{subfigure}
             \centering
             \includegraphics[width=1\linewidth]{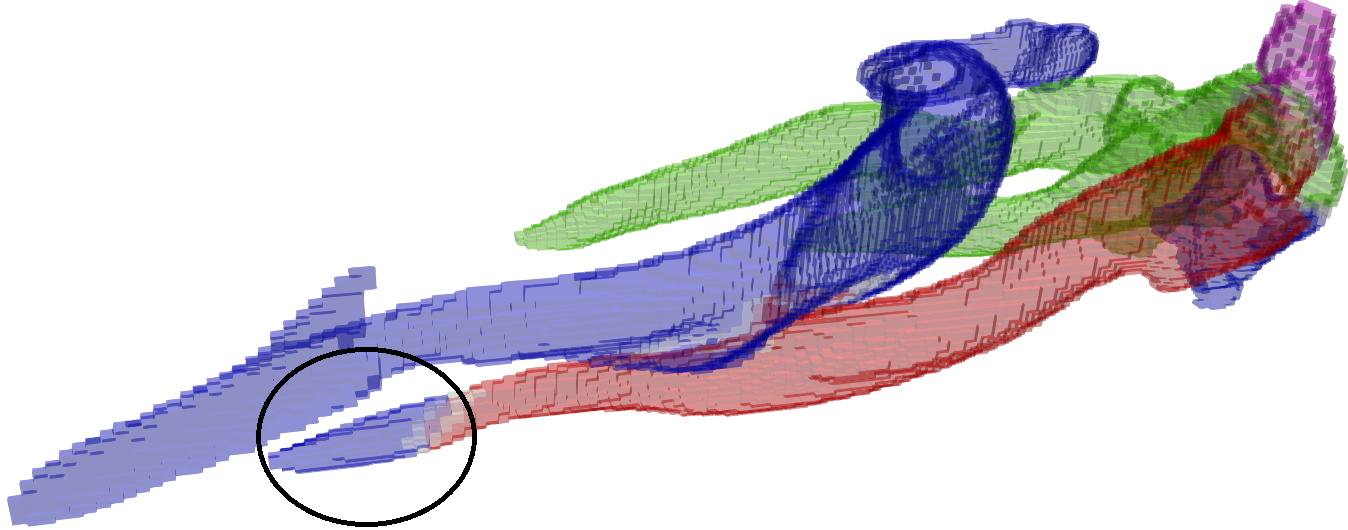}
             \vskip -10pt
             \caption*{(b)}
        \end{subfigure}
    \end{minipage}%
    \vskip -4pt
 \caption{(a) shows the vortical region (light-blue) and the underlying iso-surface components (blue, red, green, etc.). (b) shows the assigned colors to the region's cells based on the Euclidean distance from the closest iso-surface component. The highlighted area shows the wrong color (blue) assigned to the cells of the red vortex.}
 \label{fig_3_1}
\end{figure}

First, the vortex separation process described in~\cite{adeelhairpin2023} assigns IDs to cells based on their Euclidean distance from nearby isosurface components. However, this method may inaccurately assign IDs if cells are closer to the wrong iso-surface component, as depicted in \cref{fig_3_1}. Second, the separation process in~\cite{adeelhairpin2023} is streamlined by employing a histogram expansion method to select scalar values for iso-surface extraction. This involves extracting a refined histogram of the scalar field and selecting bin values corresponding to Fibonacci series indices as scalar steps. Although this approach provides a global and uniform selection of scalar values across all regions, it may result in iso-surfaces that inadequately cover vortical regions, as illustrated by the red component in \cref{fig_3_1}. Additionally, employing a global iso-value may result in insufficient or inaccurate splits, or both. %, particularly in local regions. 
Third, the amount of splitting is regulated using the Vortex Size Factor (VSF) parameter. However, determining the optimal value of VSF is not straightforward, as the size of vortices relative to the domain size can vary significantly across different flow datasets. To find an appropriate VSF value, multiple passes of the algorithm with different VSF values must be attempted, relying on the user's visual analysis for satisfaction. This iterative process can be time-consuming, particularly for large-scale datasets. To overcome the above limitations, we make the following contributions in this work.

\vspace{-0.1in}
\begin{itemize}
  \setlength{\itemsep}{0pt}
  \setlength{\parskip}{0pt}
  \itemsep0em 
    \item We present a vortex separation method that extends the contour tree-based segmentation approach~\cite{carr2003computing}. Initially, we obtain the segmentation by identifying the critical points of the contour tree. Then, employing the ``layering'' strategy (refer to \cref{sec:method}), we finalize the separation process by assigning IDs to the unsegmented cells based on the initial segmentation. 
    % Our method is an improvement over~\cite{adeelhairpin2023}.
    \item Traditional contour tree-based segmentation suffers from inaccurate splits because it only considers local critical points for segmentation, disregarding the global context of the object being segmented. To address this limitation, we compute a statistic based on the extracted vorticity lines in the vicinity of the split, determining whether a vortex should be  split at a particular level of segmentation.
\end{itemize}

To demonstrate the effectiveness, we apply our method to multiple turbulent flow datasets~\cite{li2019direct, lee2013petascale}. The results show that our approach can robustly separate entangled vortices as compared to previous methods and significantly help avoid inaccurate splits.

\section{Related Work}
Vortex identification and separation techniques for turbulent flow analysis have seen significant advancements over the years, driven by the need to better understand complex flow phenomena. Early seminal work~\cite{hunt1987vorticity} introduced the $Q$-criterion, a widely adopted method for identifying regions with high swirling motion in flows. Building upon this foundation, subsequent studies by ~\cite{chong1990general},~\cite{jeong1995identification} and~\cite{liu2018rortex} introduced $\Delta$, $\lambda_2$ and $Rortex$ criteria which furthered our understanding of vortical structures. These are \emph{region-based methods} which typically require threshold values that can impact the size and extent of the extracted vortices. 
\emph{Line-based methods}~\cite{PEI99, weinkauf2010streak, Schafhitzel2008Topology} are used to extract vortex corelines around which fluid particles revolve. Vortex coreline methods are usually parameter free, yet they often yield fragmented lines, presenting a challenge in accurately categorizing a vortex into a specific type~\cite{adeel2022hairpin}. These are local methods of vortex extraction that utilize the velocity vector at a point to calculate subsequent criteria. Additionally, global approaches such as \emph{Geometric methods}~\cite{sadarjoen2000detection}, \emph{Integration-based methods}~\cite{wiebel2011topological,weinkauf2010streak}, \emph{Objective methods}~\cite{haller2005objective,haller2015lagrangian, Sadlo2006Visualization}, and \emph{Feature level sets}~\cite{nguyen2020Taylor} offer alternative solutions. These approaches leverage streamlines, pathlines or observe the attraction behavior of injected particles over time to identify vortices.

Topological segmentation approaches based on contour-trees~\cite{carr2003computing} have been introduced to identify vortices~\cite{bremer2015identifying,schneider2008interactive,adeelhairpin2023}. ~\cite{bremer2015identifying} introduced a novel vortex detection technique based on topological analysis of a scalar indicator function. The method identifies seeds for potential vortices as local maxima/minima of the indicator function and optimizes a local threshold for each vortex using topological encoding by using a criterion called \emph{relevance}.~\cite{schneider2008interactive} presented a visualization tool facilitating the comparison of two scalar fields using iso-surfaces, extracted via the largest contour segmentation of the scalar field. In a recent study,~\cite{adeelhairpin2023} performed the contour-tree based segmentation of vortices. They first extract the vortical regions using an indicator function (e.g., $\lambda_2$) and then separate the regions using progressive extraction of iso-surfaces. This ends up in a hierarchical tree representing the split/merge relation of vortices. However, their approach presents several limitations as mentioned in \cref{sec:intro}. Vortices can undergo complex interactions, including merging, splitting, and stretching. Separating entangled vortices, particularly in turbulent flows, remains a challenging task.

% \section{Problem Statement}
% \input{content/problem_statement}

\section{Our Method}
\label{sec:method}
In this section, we first examine the topology-based vortex separation process, then discuss mitigating the inaccurate splitting issue.

\begin{figure}[t]
 \centering 
   \begin{minipage}{0.35\linewidth}
    \centering
        \begin{subfigure}
             \centering
             \includegraphics[width=1\linewidth]{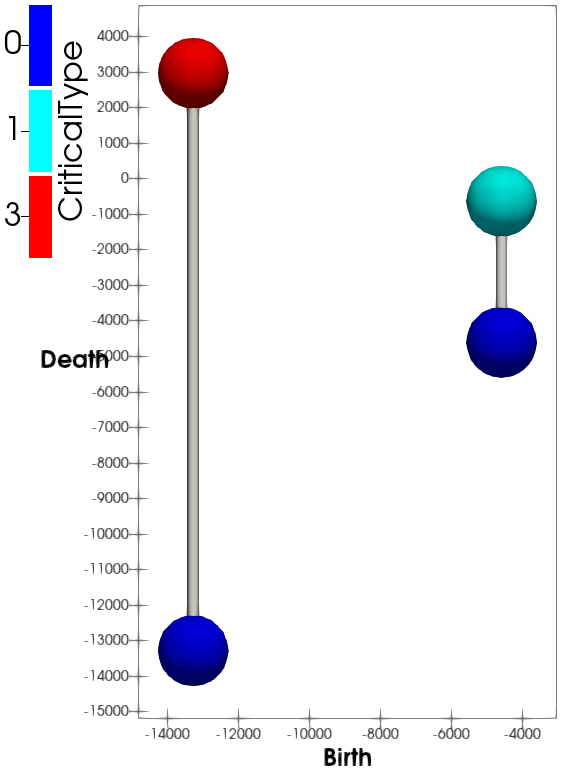}
             \vskip -8pt
             \caption*{(a)}
        \end{subfigure}
    \end{minipage}%
    \hspace{0.35in}
    \begin{minipage}{0.35\linewidth}
        \centering
        \begin{subfigure}
             \centering
             \includegraphics[width=1\linewidth]{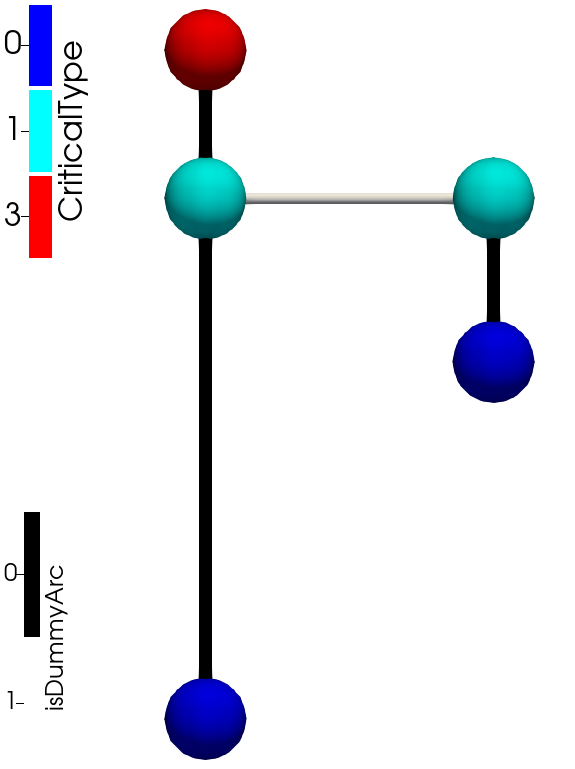}
             \vskip -8pt
             \caption*{(b)}
        \end{subfigure}
    \end{minipage}%
    \vskip -4pt
 \caption{(a) shows the persistence diagram of two critical point pairs with the highest persistence. Here maximum($\mathbb{M}$), saddle($\mathbb{S}$) and minimum($\mathfrak{m}$) points are represented by red, cyan and blue, respectively. (b) shows the corresponding minimal join tree of the chosen critical point pairs with one $\mathbb{M}$, one $\mathbb{S}$ and two $\mathfrak{m}$.}
 \label{fig_4_1}
\end{figure}

\subsection{Topology-based Vortex Separation}
\label{subsec:topology}
\begin{figure}[t]
 \centering 
      \begin{subfigure}
         \centering
         \includegraphics[width=0.95\linewidth]{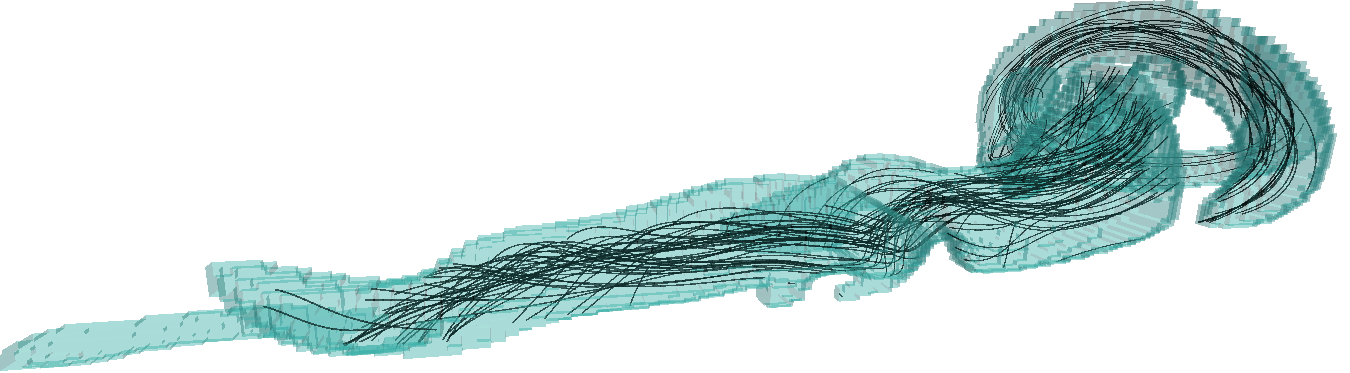}
         \vskip -15pt
         \caption*{(a)}
     \end{subfigure}
     \vskip -4pt
     \begin{subfigure}
         \centering
         \includegraphics[width=0.95\linewidth]{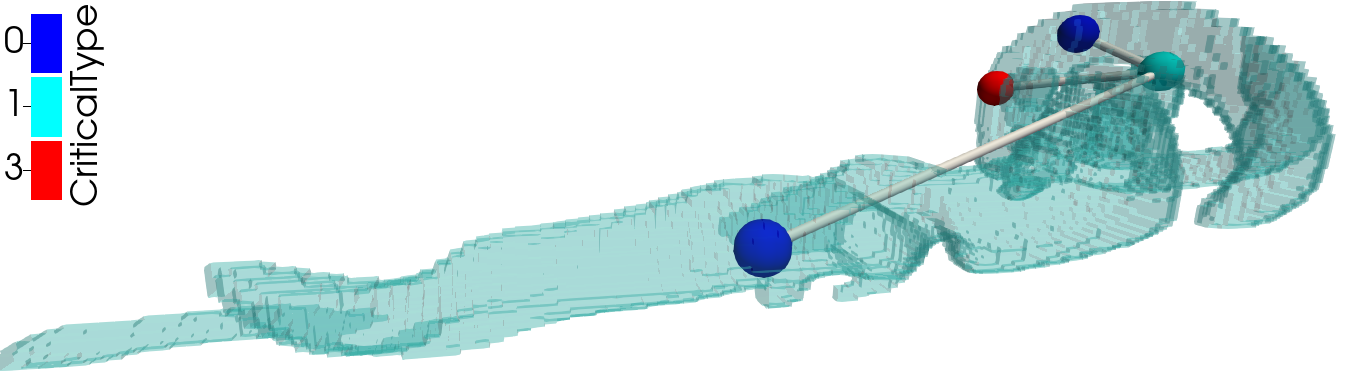}
         \vskip -15pt
         \caption*{(b)}
     \end{subfigure}
     \vskip -4pt
     \begin{subfigure}
         \centering
         \includegraphics[width=0.95\linewidth]{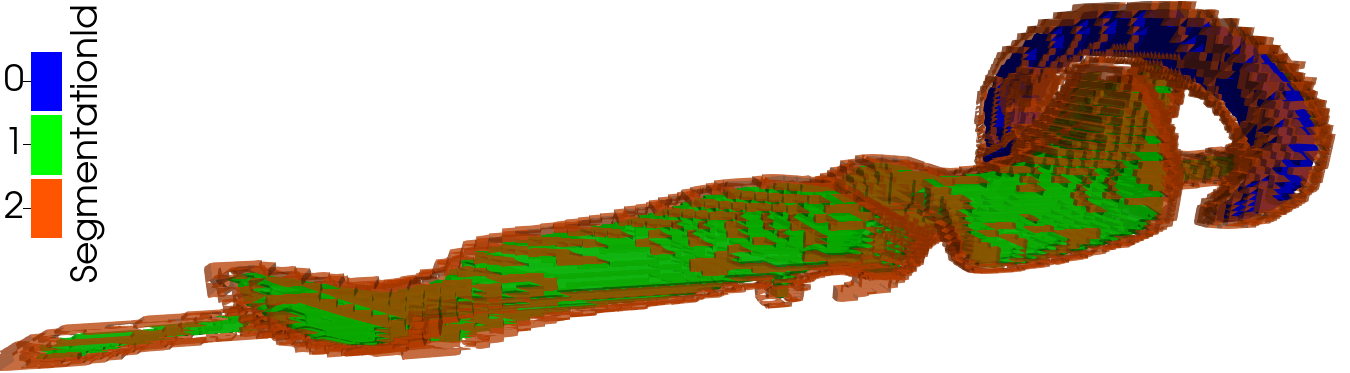}
         \vskip -15pt
         \caption*{(c)}
     \end{subfigure}
    \vskip -3pt
 \caption{(a) shows a single region containing a streamwise and a horseshoe vortex, indicated by the vorticity line (black). (b) shows the join tree embedded within the region indicating the corresponding location of the critical points. (c) shows the segmentation of the region based on the join tree, where green and blue cells correspond to two $\mathfrak{m}$-$\mathbb{S}$ pairs, and the red cells correspond to a $\mathbb{M}$-$\mathbb{S}$ pair.}
 \label{fig_4_2}
\end{figure}

Contour trees~\cite{carr2003computing} encode the merging and splitting relations of the level sets of scalar fields. For a set of points in 3D space ${p \in \mathbf{R}^3}$ and a scalar field ${f(p) \in \mathbf{R}}$, a level set is defined as $\{p \in \mathbf{R}^3 \ | \ f(p) = c\}$. As the value of $\textit{c}$ changes, the level sets evolve, splitting and merging, which is encoded by split and join trees, respectively. The evolution of level sets occurs exclusively at topological critical points where $\nabla f(p) = \mathbf{0}$. In the case of a join tree, the leaf nodes represent the minima ($\mathfrak{m}$) of $f(p)$, and as we increase the value of $f(p)$ the level sets merge at the saddle ($\mathbb{S}$) points of $f(p)$. Assuming a region with a simplified scalar field having only one $\mathbb{S}$, the smallest possible join tree can be represented as shown in \cref{fig_4_1}(b). The region can be segmented by assigning IDs to points ${p}$ based on their scalar value ${f(p)}$ falling within one of three ranges corresponding to the three edge pairs of the join tree. The three edge pairs are ($\mathbb{S}$-$\mathbb{M}$), ($\mathfrak{m}_1$-$\mathbb{S}$) and ($\mathfrak{m}_2$-$\mathbb{S}$), where $\mathbb{M}$ = Maximum. The segment seeds from the first critical point of the pair and continues to grow until the second critical point is reached, making the segments distinct from each other. This segmentation is depicted in \cref{fig_4_2}(c) using red, blue, and green colors, respectively. 

%Mathematically

%\begin{equation}
% S(p) = \begin{cases} 
%      0 & min_1\leq f(p)\leq sad \\
%      1 & min_2\leq f(p)\leq sad \\
%      -1 & sad\leq f(p)\leq max 
%   \end{cases}
%\end{equation}

%where $S(p)$ denotes the segmentation ID assigned to the point $p$, $f(p)$ is the scalar value, and $min_1$, $min_2$, $sad$ are the values of Minimum$_1$, Minimum$_2$, and Saddle points respectively.

% \begin{figure}[t]
%  \centering
%  \includegraphics[width=1\linewidth]{figures/fig4_1.png}
%  \vskip -5pt
%  \caption{(a) shows a join tree featuring a maximum, a saddle, and two minima. (b) illustrates a single region containing a streamline and a horseshoe vortex, indicated by the vorticity line (in black). (c) displays the join tree embedded within the region and indicates the corresponding location of the critical points. (d) showcases the segmentation of the region based on the join tree, where green and blue cells correspond to two minima-saddle pairs, and the red cells correspond to a maximum-saddle pair.}
%  \label{fig_4_1}
% \end{figure}

\begin{figure}[t]
 \centering 
      \begin{subfigure}
         \centering
         \includegraphics[width=0.95\linewidth]{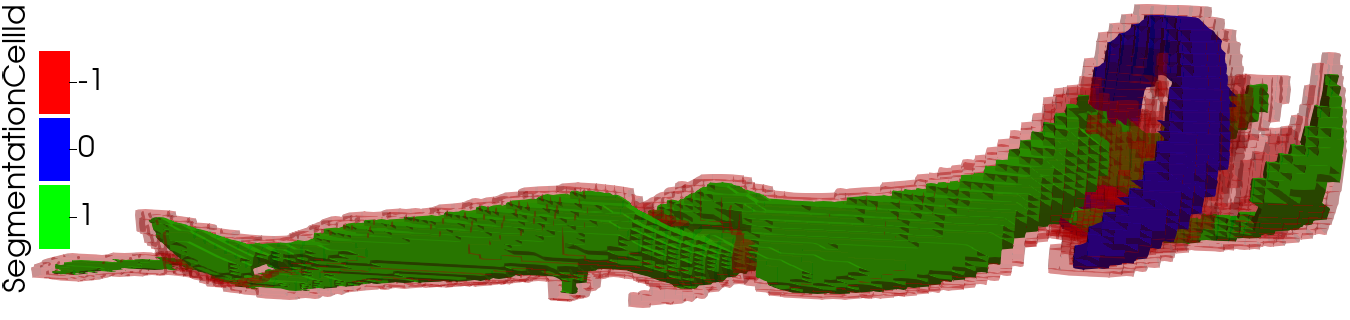}
         \vskip -7pt
         \caption*{(a)}
     \end{subfigure}
     \vskip -5pt
     \begin{subfigure}
         \centering
         \includegraphics[width=0.95\linewidth]{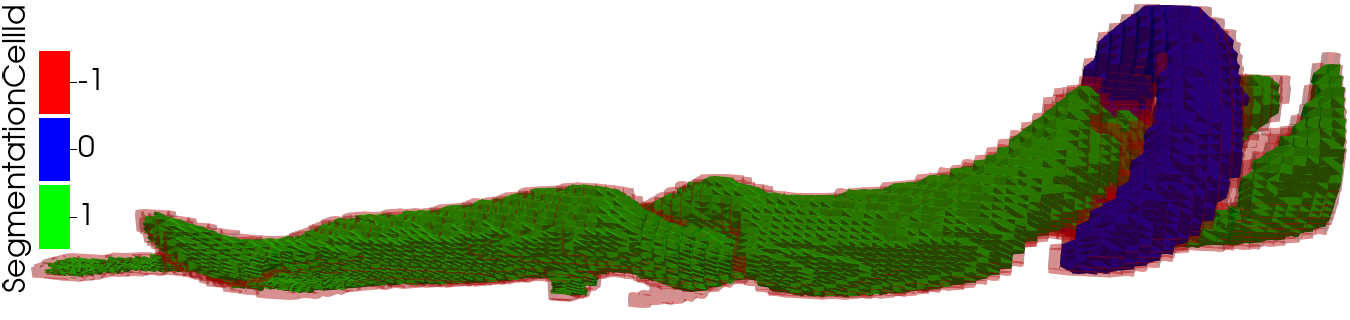}
         \vskip -7pt
         \caption*{(b)}
     \end{subfigure}
     \vskip -5pt
     \begin{subfigure}
         \centering
         \includegraphics[width=0.95\linewidth]{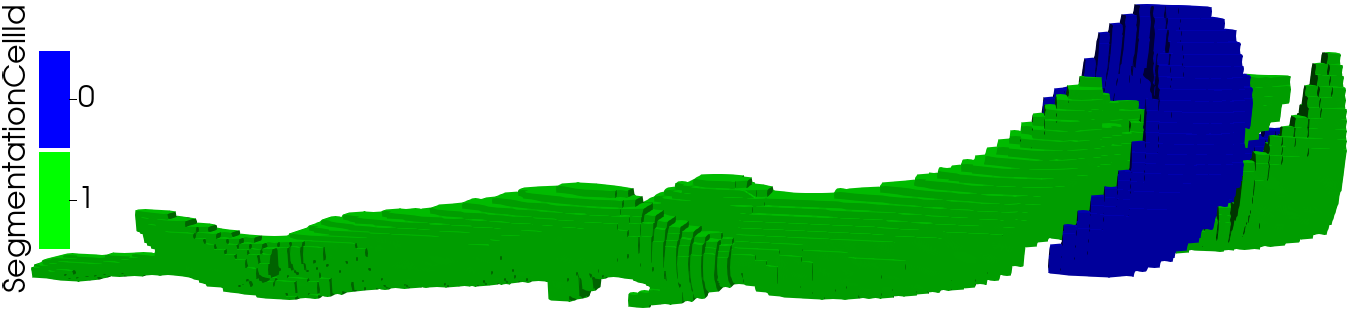}
         \vskip -7pt
         \caption*{(c)}
     \end{subfigure}
     \vskip -5pt
 \caption{(a) shows the initial segments obtained from the minimal join tree. IDs $0$ and $1$ represent the seed segments corresponding to the ($\mathfrak{m}_1$-$\mathbb{S}$) and ($\mathfrak{m}_2$-$\mathbb{S}$) pairs, respectively. $-1$ is the ID of the query segment corresponding to the ($\mathbb{S}$-$\mathbb{M}$) pair. (b) displays the same region after several iterations of the layering process. It is evident that the seed segments have expanded, resulting in fewer cells remaining in the query segment. (c) demonstrates the region completely separated at the conclusion of the layering process.}
 \label{fig_4_3}
 \end{figure}

Given a minimal join tree, our goal is to separate the region into exactly two vortices. The segments corresponding to the $\mathfrak{m}$-$\mathbb{S}$ pairs belong to the vortices that are to be separated (\textit{blue and green in \cref{fig_4_2}(c)}). We call them the seed segments. To achieve complete separation of the vortical region into exactly two vortices, one of the IDs from the seed segments must be assigned to the segment corresponding to the $\mathbb{M}$-$\mathbb{S}$ pair (\textit{red in \cref{fig_4_2}(c)}). We call this the query segment. While a strategy similar to~\cite{adeelhairpin2023} could be employed to assign IDs to the query segment based on the Euclidean distance to the seed segment, it may suffer from the limitations discussed in \cref{sec:intro}. To rectify this limitation, an accurate measure of distance is required, such as graph geodesic distance. However, employing graph geodesic distance in this scenario is computationally demanding, as it requires computing distances between each cell in the query segment and all cells in the seed segment to find the nearest seed segment. This is where our ``layering'' strategy comes in. Layering is visualized in \cref{fig_4_3} and works as follows:
\begin{enumerate}
\itemsep0em 
\item Identify cells in the query segment that are immediate neighbors of the cells in the seed segments. We refer to this collection of cells as a \textit{layer}.
\item Assign IDs to the cells in the \textit{layer} based on the ID of the closest neighboring cell in the seed segments.
\item Iterate through steps (1) and (2) until there are no cells remaining in the query segment.
\end{enumerate}

Critical point pairs of the minimal join tree are chosen based on the persistence~\cite{edelsbrunner2022computational} of the pairs. For this purpose, we get the persistence diagram of the scalar field and choose a ($\mathbb{M}$-$\mathfrak{m}$) and a ($\mathbb{S}$-$\mathfrak{m}$) pair with the highest persistence as depicted in \cref{fig_4_1}(a). \changed{In this paper, we use $\lambda_2$-criterion as the scalar field but the method is equally valid for other region-based criterion such as $Q$\cite{hunt1987vorticity}, $\Delta$\cite{jeong1995identification}, $OW$\cite{okubo1970horizontal, weiss1991dynamics}, etc.}. Initially, we extract the vortical regions utilizing the region growing strategy from~\cite{adeelhairpin2023} using $\lambda_2$ criterion, then our vortex separation approach works as follows. For each disconnected region, we do the following:

\begin{enumerate}
\itemsep0em 
    \item Get persistence diagram of the input scalar field for the region.
    \item Choose critical point pairs as depicted in \cref{fig_4_1}(a) with the highest persistence.
    \item Simplify the topology by removing all remaining critical points in the region based on the selected critical points.
    \item Extract the join tree and get the initial segmentation in the form of seeds and query segments.
    \item If both seed segments have at least one cell, then do ``layering'' and split the region. Otherwise, go to (2) and pick a new $\mathfrak{m}$-$\mathbb{S}$ pair with lower persistence. \changed{Stop, if no $\mathfrak{m}$-$\mathbb{S}$ pairs are left}.
    \item \changed{Automatically check whether to avoid the split using \cref{eq1}}. If it needs to be avoided, go to step (2) and pick a new $\mathfrak{m}$-$\mathbb{S}$ pair with lower persistence. Otherwise, finalize the split and record the changes in the vortex hierarchy.
    \item For each new region, do (1)--(6) recursively.
\end{enumerate}

\subsection{Check and Avoid Inaccurate Splits}
\label{sec:avoid_splits}
\begin{figure}[t]
 \centering 
   \begin{minipage}{0.5\linewidth}
    \centering
        \begin{subfigure}
             \centering
             \includegraphics[width=1\linewidth]{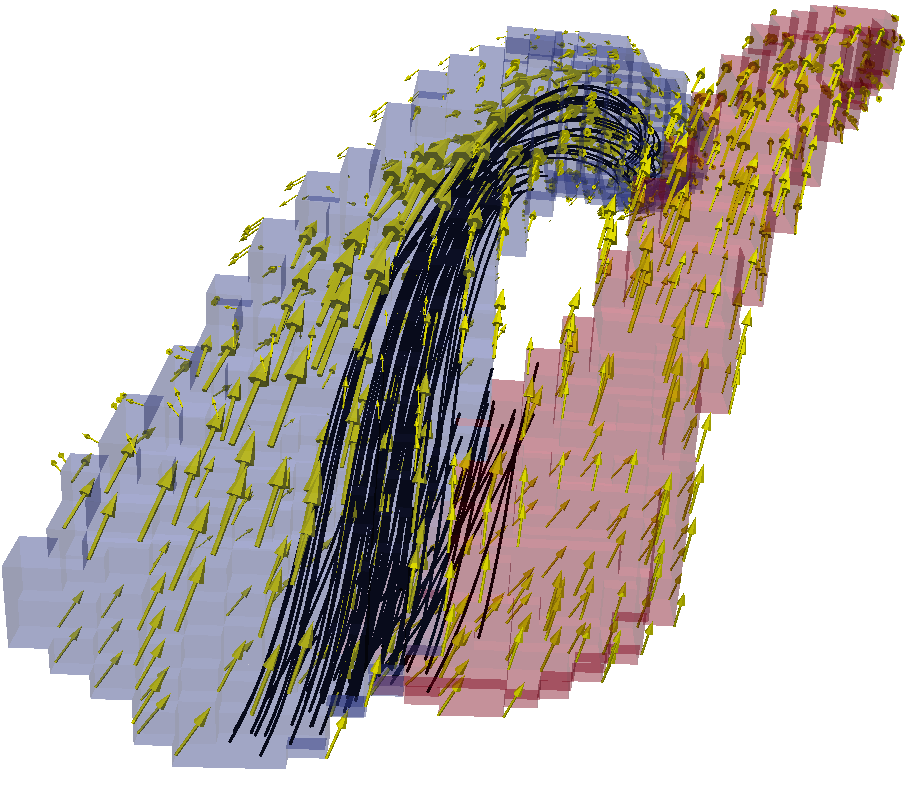}
             \vskip -10pt
             \caption*{(a)}
        \end{subfigure}
    \end{minipage}%
    \begin{minipage}{0.5\linewidth}
        \centering
        \begin{subfigure}
             \centering
             \includegraphics[width=1\linewidth]{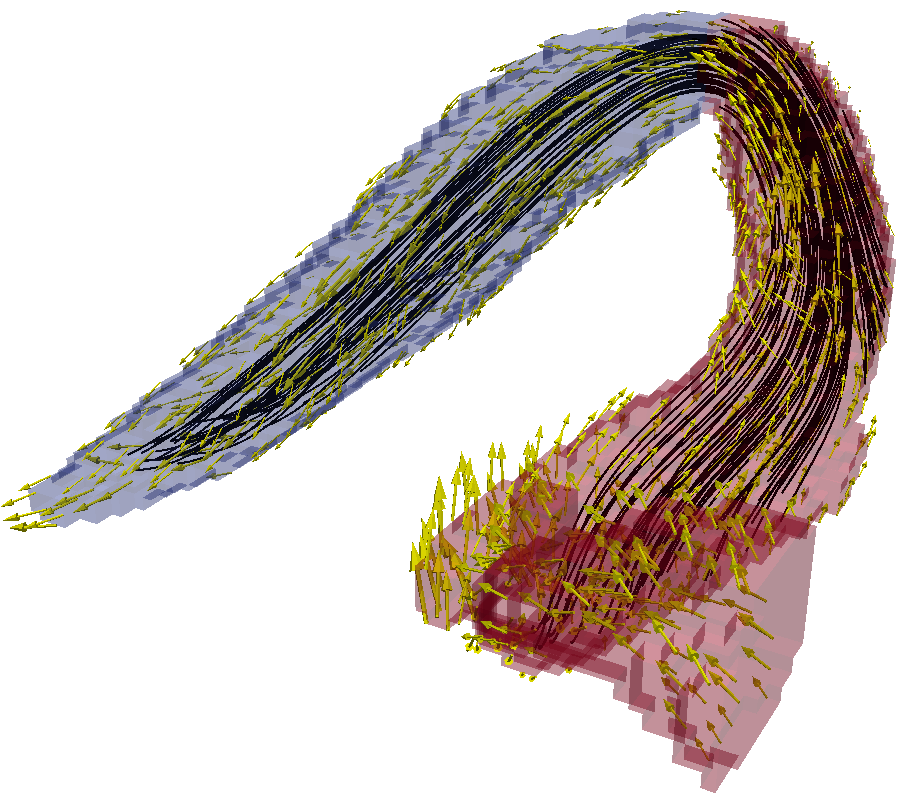}
             \vskip -10pt
             \caption*{(b)}
        \end{subfigure}
    \end{minipage}%
    \vskip -4pt
 \caption{This figure shows how the vorticity lines help avoid the inaccurate split problem. (a) shows a valid split of two vortices (blue and red) as $R_1$ value of the cells overlapped by the vorticity lines (black) is close to 0. (b) shows an inaccurate split where the $R_1$ value is close to 1 which subsequently is avoided. Arrows (yellow) show the direction of the $vorticity$ vectors.}
 \label{fig_4_4}
\end{figure}

To avoid inaccurate splits, we extract and utilize vorticity lines close to the boundary of the split. We move a few layers of cells away from the boundary towards the larger region and uniformly select $min(100, \# \ of\ points\ in\ the\ bigger\ region)$ as seeds for the vorticity lines. \changed{Our algorithm checks and decides whether the split should be avoided based on the following equation},
\begin{equation}
R_1 \begin{cases} 
     < 0.25 & \rightarrow \ $Split$ \; , \\
     \geq 0.25 & \rightarrow \ $!Split$ \; , \\
     % > 0.1 < 0.25 & \rightarrow if\ \frac{R_1}{R_2} \leq 0.25 \rightarrow $Split$ \ else \ $!Split$\\
  \end{cases}
  \label{eq1}
\end{equation}

\noindent where $R_1$ represents the percentage of cells overlapped by the vorticity lines in the smaller region. This examines if the vorticity lines' trend persists across the split boundary. If a large area is covered by the vorticity lines in the smaller region originating from the larger region, the trend is consistent, and the region should not be split, as shown in \cref{fig_4_4}(b). The reason behind selecting seeds close to, rather than precisely at, the boundary is exemplified in \cref{fig_4_4}(a). In this scenario, two vortices form a "V" shape as indicated by the arrows of the $vorticity$ vectors. If the boundary points were utilized as seeds, the vorticity lines could extend into both regions, resulting in a high value of $R_1$ despite being an inaccurate split case. This discrepancy arises because the boundary doesn't precisely delineate vortices but rather marks the topological boundaries of vortical regions identified by contour tree-based segmentation and layering. We shift 5 layers of cells away from the boundary points towards the larger region and utilize their points as the seeds \textit{(more details in the supplemental)}. In the earlier levels of splitting, the region sizes are bigger and the boundary could have multiple interfaces. We let such regions split because the physics of the vorticity lines through multiple interfaces of the boundary is complex and a simple value such as $R_1$ doesn't suffice. Therefore we only use this strategy when the boundary has only one interface. \changed{Additionally, vorticity lines may not follow the vortex shape and may terminate prematurely, especially in weaker vortices. Our method fails to avoid splits in these cases. Thus, while our method significantly improves split accuracy, it cannot completely prevent inaccurate splits.} We leave such complex scenarios for future work.

\section{Results}
\label{sec:results}
% In the following, we first discuss how our algorithm overcomes the limitations \textbf{1}, \textbf{2} and \textbf{3}. Then we apply our method to turbulent flow datasets and showcase several instances where our algorithm demonstrates better results as compared to the previous approach~\cite{adeelhairpin2023}.
Our method avoids the wrong ID assignment problem by employing the layering strategy (\cref{subsec:topology}). The seed segments expand uniformly until the region is completely separated (\cref{fig_4_3}), which indirectly approximates the graph geodesic distance. Critical points are leveraged for the separation process, where scalar values dynamically adapt to the current region, unlike the global thresholds utilized in the previous approach~\cite{adeelhairpin2023}. By using critical points for separation, we address the issue of regions being inadequately covered by iso-surfaces or not being sufficiently split. Furthermore, we utilize seed segments instead of iso-surfaces for ID assignment, eliminating the need for an additional processing step to filter unnecessary iso-surface components. Finally, our vortex separation approach is adaptive. At each level, the region is split into exactly two segments based on the minimal join tree. The process continues until the specified stop condition is met. It eliminates the necessity to experiment with multiple values of VSF or rely on visual cues from the user. The only user-specified parameter required is the value of $R_1$ in \cref{eq1} which does not introduce any of the issues outlined in \cref{sec:intro}. In the following, we apply our method to turbulent flow datasets and showcase several instances where our algorithm demonstrates better results as compared to the previous approach~\cite{adeelhairpin2023}.

\subsection{Vortex Separation Results}
% In this section, we compare our results with those obtained using the previous approach~\cite{adeelhairpin2023} to illustrate the superior performance of our vortex separation method. We begin by presenting results obtained from the stress-driven turbulent Couette flow dataset~\cite{li2019direct}. Specifically, we focus on areas characterized by the ejection of low-speed fluid flow from the boundary, as outlined in Sec. III(D) of~\cite{li2019direct}. As depicted in \cref{fig_5_1}, our method not only effectively splits the vortices but also significantly addresses the inaccurate split problem. \cref{fig_5_1}(d) and \cref{fig_5_1}(f) show the one-legged hairpin vortex carved out by our method and by~\cite{adeelhairpin2023}, respectively. Notably, \cref{fig_5_1}(f) highlights additional regions still attached to the hairpin vortex, as evidenced by the vorticity lines showing no trend continuation into those regions. In contrast, our method accurately removes those extra regions, especially the banana-shaped vortex which forms a loop. Moreover, as observed in \cref{fig_5_1}(c) and \cref{fig_5_1}(d), our method notably avoids evident inaccurate splits. Although some inaccurate splits still occur, especially at the edges of the vortices as highlighted in \cref{fig_5_1}(c), they do not pose a risk of misclassifying a vortex and can be regarded as noise. For instance, as demonstrated in \cref{fig_4_4}(b), when a hairpin vortex splits inaccurately, it forms two streamwise vortices, hindering subsequent analysis aimed at identifying specific vortex populations within the flow.

\begin{figure}[!t]
 \centering 
   \begin{minipage}{0.45\linewidth}
    \centering
        \begin{subfigure}
             \centering
             \includegraphics[width=1\linewidth]{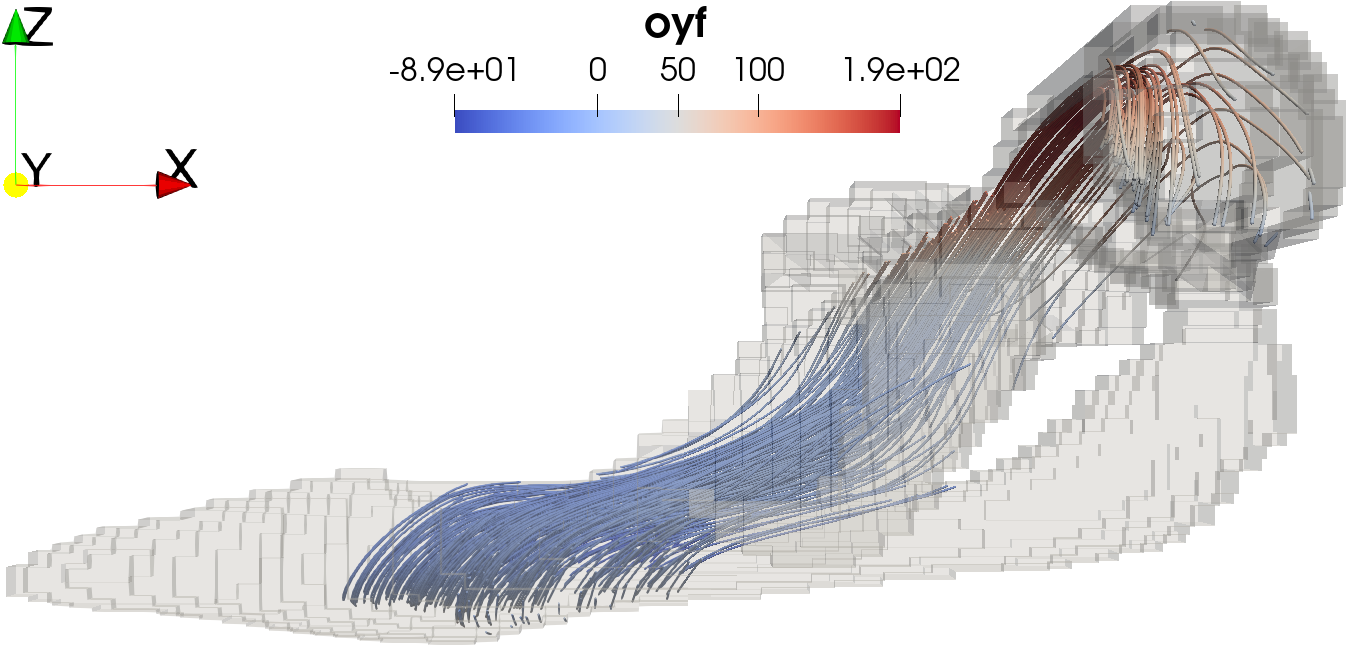}
             \vskip -5pt
             \caption*{(a)}
        \end{subfigure}
    \end{minipage}%
    \begin{minipage}{0.45\linewidth}
        \centering
        \begin{subfigure}
             \centering
             \includegraphics[width=1\linewidth]{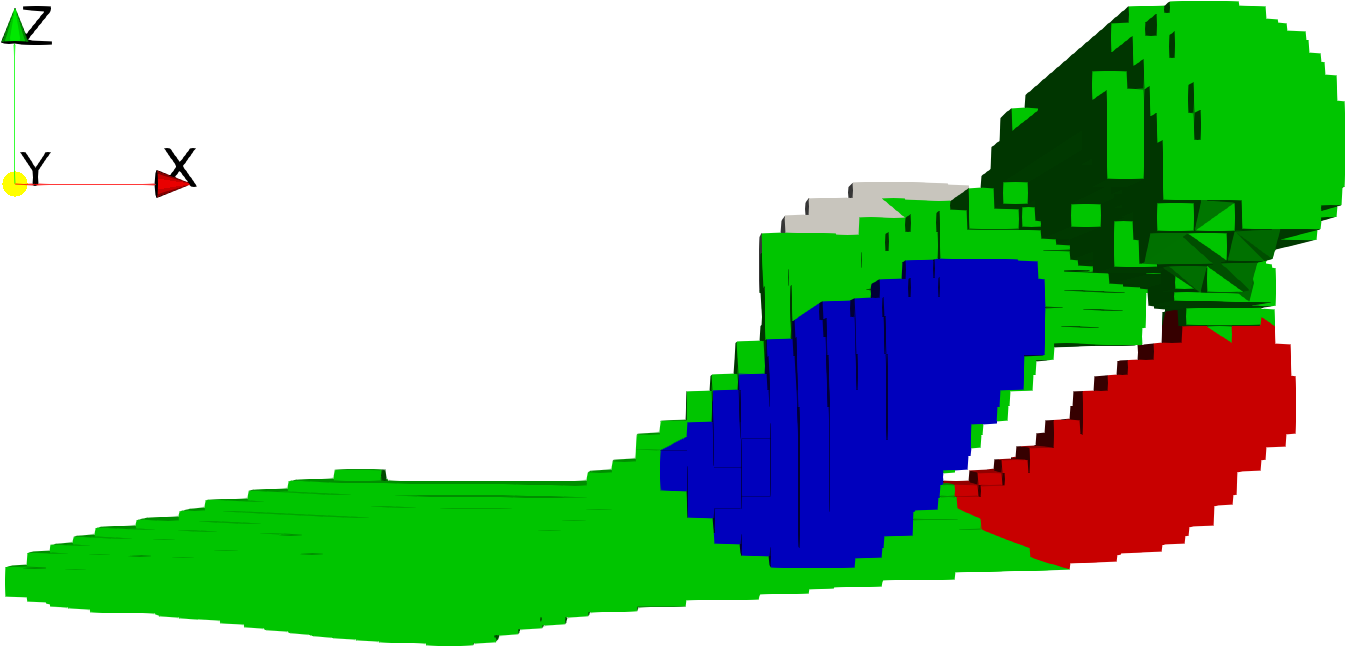}
             \vskip -5pt
             \caption*{(b)}
        \end{subfigure}
    \end{minipage}%
    \vskip 2pt
   \begin{minipage}{0.45\linewidth}
    \centering
        \begin{subfigure}
             \centering
             \includegraphics[width=1\linewidth]{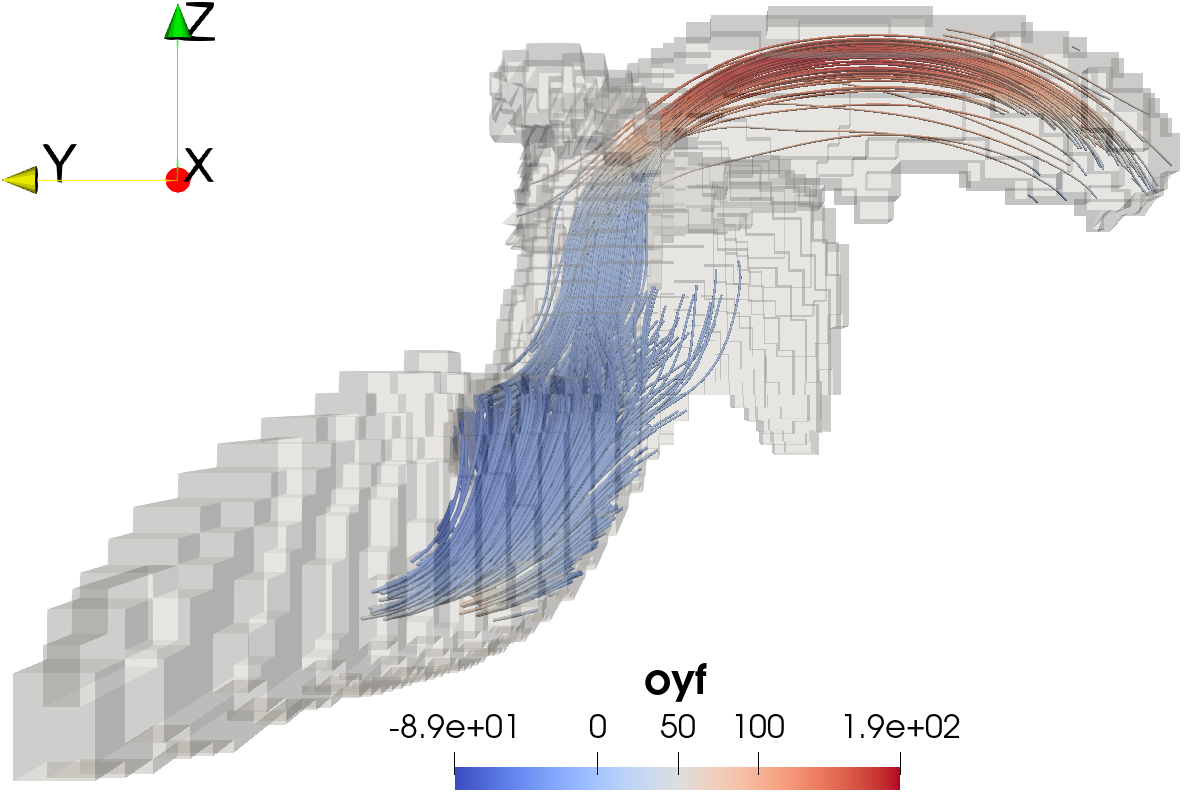}
             \vskip -5pt
             \caption*{(c)}
        \end{subfigure}
    \end{minipage}%
    \begin{minipage}{0.45\linewidth}
        \centering
        \begin{subfigure}
             \centering
             \includegraphics[width=1\linewidth]{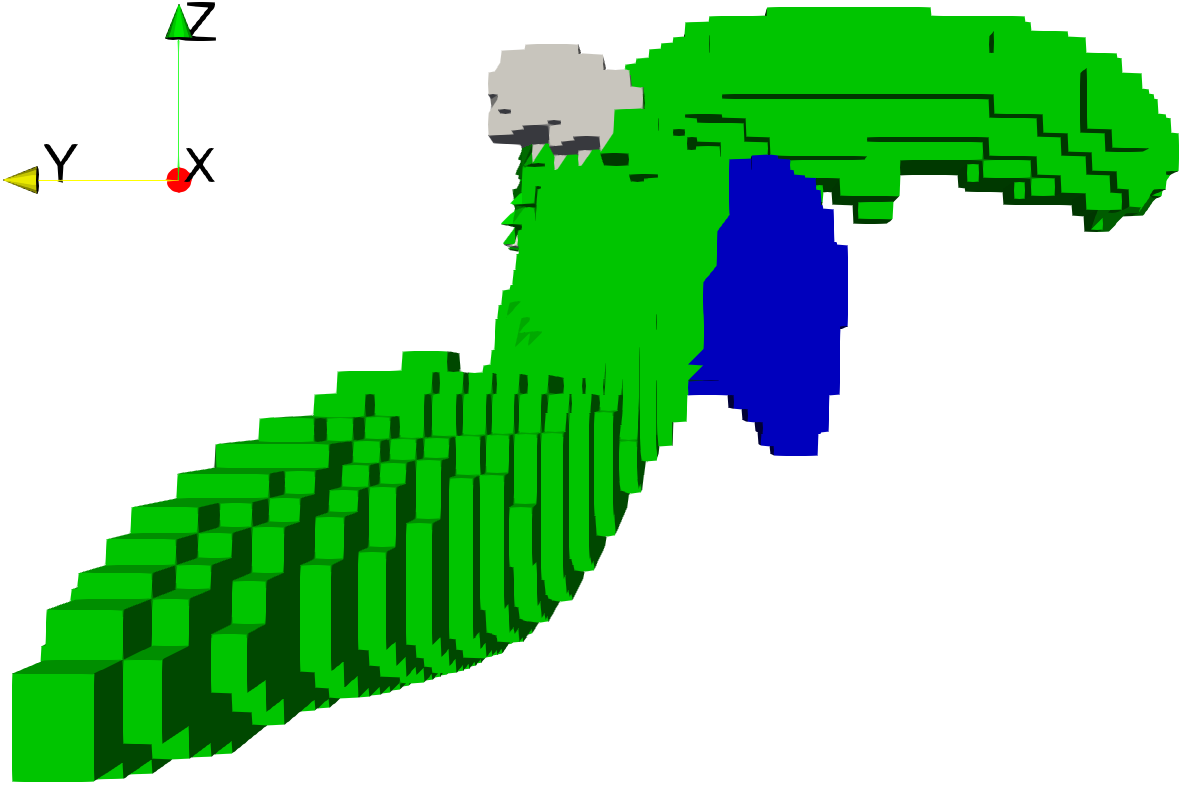}
             \vskip -5pt
             \caption*{(d)}
        \end{subfigure}
    \end{minipage}%
 \caption{(a) illustrates a hairpin vortex, evident from the strong positive and strong negative values of $\omega_y'$ in the head and leg of the hairpin vortex, respectively, as indicated by the color of the vorticity lines. This represents the finalized separation using the method in~\cite{adeelhairpin2023}, denoted by the single color of the region. (b) depicts the same vortex with our method, showcasing individual separated segments in different colors. (c) presents the same vortex from a different angle, while (d) displays our results from the same angle as (c). It is evident that our vortex extraction method removes extra blobs (\textit{blue, red, white}) while retaining the vortex of interest (\textit{green}).}
 \label{fig_5_3}
\end{figure}

In this section, we compare our results with the previous approach~\cite{adeelhairpin2023} to illustrate the superiority of our vortex separation method. We concentrate on three primary areas of improvement, namely: (1) \textbf{Adaptive Splits}, (2) \textbf{Avoiding Inaccurate Splits}, and (3) \textbf{Ensuring Sufficient Splits}. \cref{fig_5_3} shows an example of a (one-legged) hairpin vortex from the Couette flow dataset~\cite{li2019direct}. This hairpin vortex is a part of a cluster of vortices (\cref{fig_3_1}) found in the vicinity of a low-speed streak as mentioned in Sec. III(D) of~\cite{li2019direct}.  It is evident from the trend of the vorticity lines in \cref{fig_5_3} that our method effectively removes unwanted segments (\textit{blue, red, and white} in \cref{fig_5_3}(b, d)) while preserving the desired segment (\textit{green} in \cref{fig_5_3}(b, d)). This success is attributed to our utilization of local critical points for separation, as opposed to the global thresholds extracted using a statistical method, as was done in~\cite{adeelhairpin2023}, which failed to achieve the required separation, as illustrated in \cref{fig_5_3}(a). 

\begin{figure}[!t]
 \centering 
   \begin{minipage}{0.5\linewidth}
    \centering
        \begin{subfigure}
             \centering
             \includegraphics[width=1\linewidth]{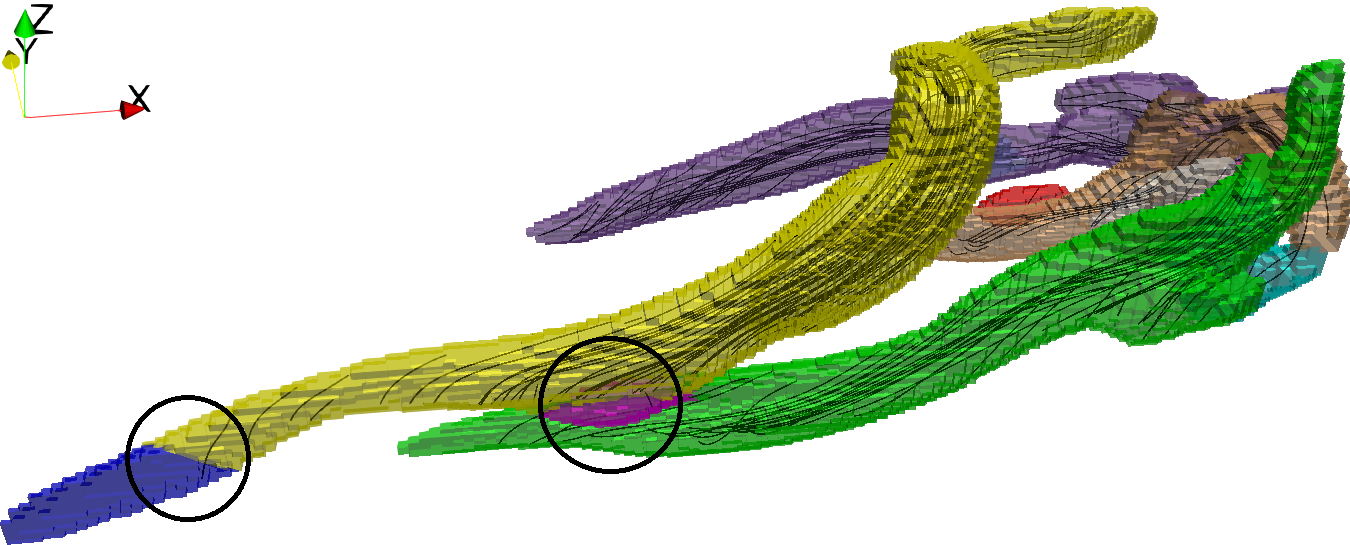}
             \vskip -10pt
             \caption*{(a)}
        \end{subfigure}
    \end{minipage}%
    \begin{minipage}{0.5\linewidth}
        \centering
        \begin{subfigure}
             \centering
             \includegraphics[width=1\linewidth]{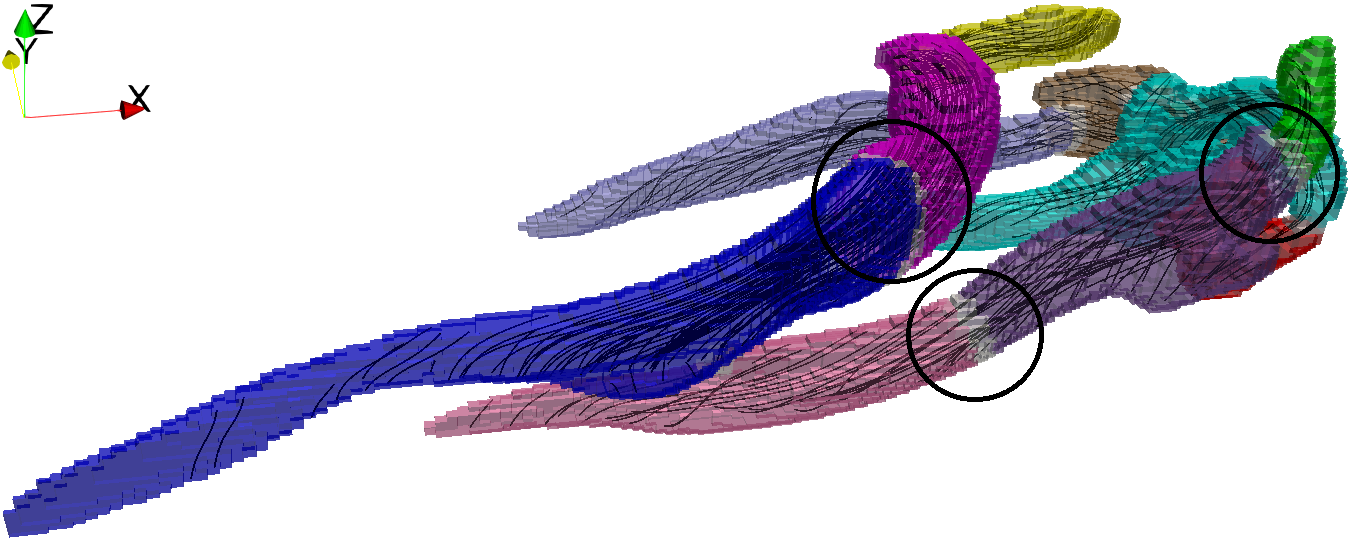}
             \vskip -10pt
             \caption*{(b)}
        \end{subfigure}
    \end{minipage}%
 \caption{(a) shows the finalized separation of a cluster of vortices using our method. Individual vortices are represented by different colors. (b) shows the finalized separation using the method in~\cite{adeelhairpin2023} where inaccurate splits are highlighted in circles. It can be clearly seen that our method demonstrates better results in avoiding inaccurate splits.}
 \label{fig_5_4}
\end{figure}

We compare the inaccurate split results in \cref{fig_5_4}. Some obvious splits are highlighted in \cref{fig_5_4}(b) where vorticity lines show a clear trend continuation. Our method avoids such splits as shown in \cref{fig_5_4}(a). Although some inaccurate splits still occur, especially at the edges of the vortices as highlighted in \cref{fig_5_4}(a), they do not pose a risk of misclassifying a vortex and can be regarded as noise. For instance, as demonstrated in \cref{fig_4_4}(b), when a hairpin vortex splits inaccurately, it forms two streamwise vortices, hindering subsequent analysis aimed at identifying specific vortex populations within the flow. Furthermore, these inaccurate segments appear in regions where the strength of vorticity is notably low \changed{and the vorticity lines integrate out of the domain without following the shape of the vortex (\textit{blue vortex} in \cref{fig_5_4}(a))}.

\begin{figure}[t]
 \centering 
   \begin{minipage}{1\linewidth}
    \centering
        \begin{subfigure}
             \centering
             \includegraphics[width=1\linewidth]{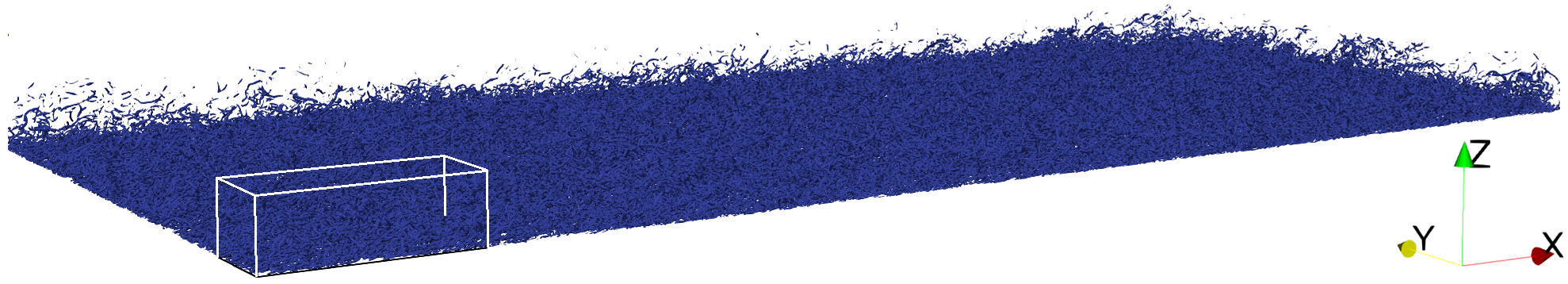}
             \vskip -10pt
             \caption*{(a)}
        \end{subfigure}
    \end{minipage}%
    \vskip 0.1pt
    \begin{minipage}{0.5\linewidth}
    \centering
        \begin{subfigure}
             \centering
             \includegraphics[width=1\linewidth]{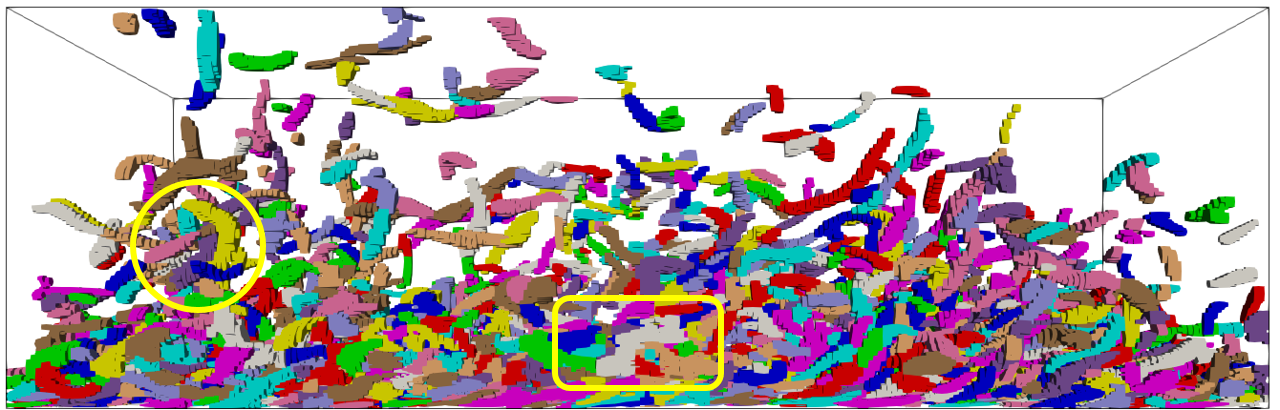}
             \vskip -5pt
             \caption*{(b)}
        \end{subfigure}
    \end{minipage}%
    \begin{minipage}{0.5\linewidth}
    \centering
        \begin{subfigure}
             \centering
             \includegraphics[width=1\linewidth]{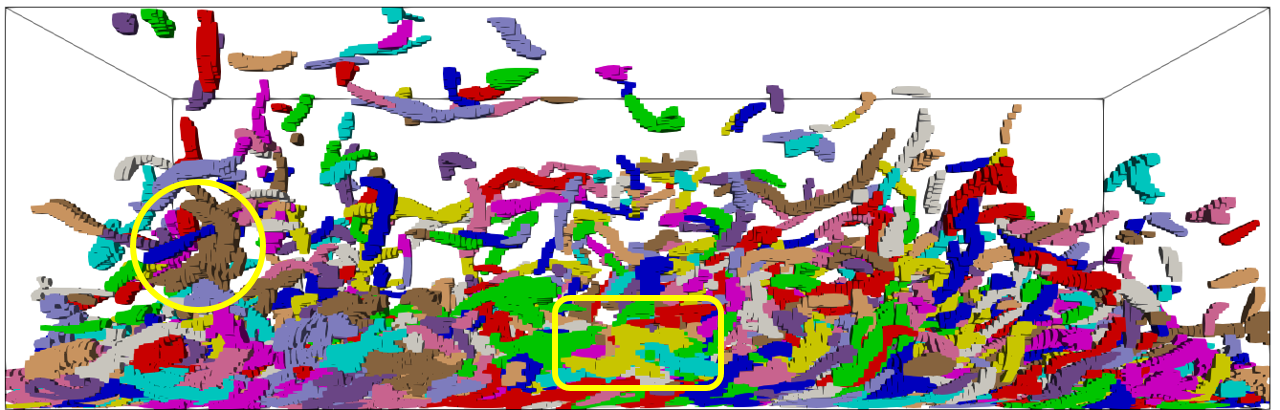}
             \vskip -5pt
             \caption*{(c)}
        \end{subfigure}
    \end{minipage}%
    \vskip 0.1pt
    \begin{minipage}{0.5\linewidth}
    \centering
        \begin{subfigure}
             \centering
             \includegraphics[width=1\linewidth]{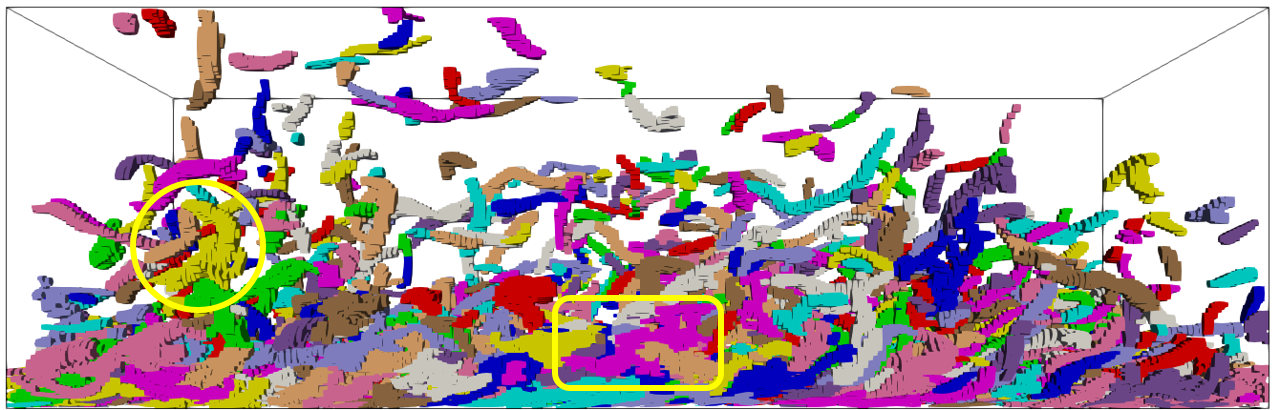}
             \vskip -5pt
             \caption*{(d)}
        \end{subfigure}
    \end{minipage}%
    \begin{minipage}{0.5\linewidth}
    \centering
        \begin{subfigure}
             \centering
             \includegraphics[width=1\linewidth]{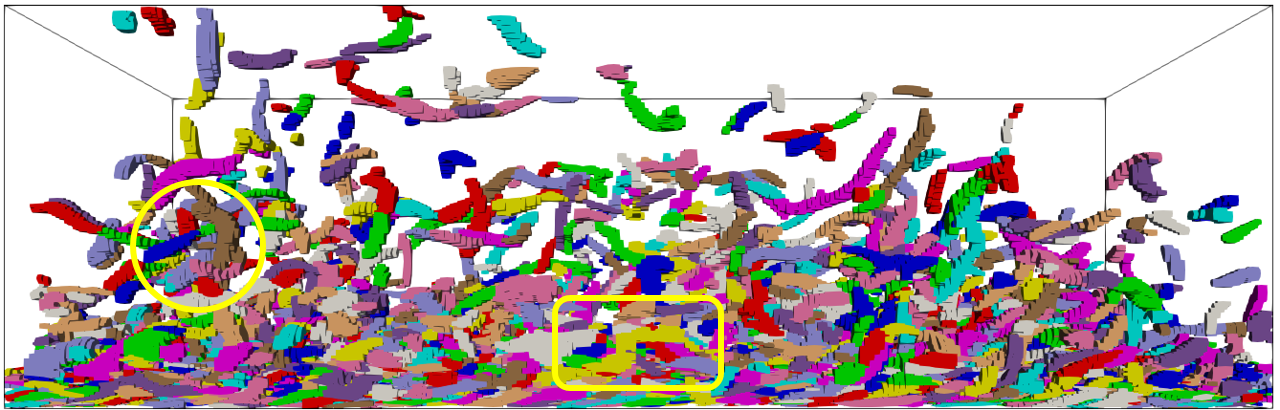}
             \vskip -5pt
             \caption*{(e)}
        \end{subfigure}
    \end{minipage}%
 \caption{(a) shows the vortices (blue) in a single timestamp of the dataset from~\cite{lee2013petascale}. For effective visualization, we only show the results for the section highlighted in a rectangle (white) in (a). Fig. (b), (c), and (d) show the separation results for~\cite{adeelhairpin2023} for VSF values of 1, 3, and 5 respectively. (e) shows the results of our separation method. Individual vortices are visualized with different colors.}
 \label{fig_5_5}
\end{figure}

In \cref{fig_5_5}, we share the results on $1/64$-th subset of a single timestamp of the turbulent channel flow at a friction Reynolds number $Re_\tau\sim1000$ based on Johns Hopkins Turbulence Database (JHTDB)~\cite{lee2013petascale} as shown in \cref{fig_5_5}(a). It can be seen in \cref{fig_5_5}(b--d) that the separation results significantly differ based on different values of VSF parameter of~\cite{adeelhairpin2023}. Some vortices are split insufficiently while others are inaccurately split as shown in the highlighted areas. It is hard to determine even with the user's visual analysis which split should be considered accurate. In contrast, \changed{our separation method is regulated by a less sensitive parameter ($R_1$)}, making our results more uniform and robust as shown in \cref{fig_5_5}(e).

\section{Discussion and Future Work}
In this work, we presented a vortex separation technique based on scalar field critical points. In order to overcome several limitations of the previous method, we introduced the ``layering'' strategy and partially overcame the inaccurate split problem with the statistical inclusion of vorticity lines. Our algorithm's performance is slower compared to that of~\cite{adeelhairpin2023}, attributable to two main factors. Firstly, our method explores all critical point pairs within the underlying region for separation, which could be considerably higher in number compared to a global threshold selection approach. Secondly, the layering strategy introduces additional processing time. This trade-off between performance and accuracy suggests potential improvements by implementing criteria such as persistence to curtail the number of critical points considered. We also plan to further explore the interactions of vorticity lines when the split boundary is relatively complex having multiple interfaces. %Furthermore, a criterion to cancel ineffective critical points specific to vortices will be explored.

\section*{Supplemental Materials}
\label{sec:supplemental_materials}
The supplemental document contains the pseudo-code, missing details, additional results, and performance analysis of the method.

\acknowledgments{
This research was supported by NSF OAC 2102761.}

\bibliographystyle{abbrv-doi-hyperref-narrow}

\bibliography{main}

%\clearpage
%\appendix % You can use the `hideappendix` class option to skip everything after \appendix
%\input{content/supplemental}

\end{document}

\setlength{\abovedisplayskip}{1pt}
\setlength{\belowdisplayskip}{1pt}
\setlength{\abovedisplayshortskip}{3pt}
\setlength{\belowdisplayshortskip}{-2pt}
\setlength{\belowcaptionskip}{3pt}
\setlength{\abovecaptionskip}{2pt}
\setlength{\textfloatsep}{9pt}
\setlength{\floatsep}{3pt}
\setlength{\intextsep}{2pt}

\maketitle

\setcounter{section}{0}
\setcounter{figure}{0}
\setcounter{table}{0}
\setcounter{equation}{0}

\begin{figure}[!t]
 \centering 
   \begin{minipage}{0.48\linewidth}
    \centering
        \begin{subfigure}
             \centering
             \includegraphics[width=1\linewidth]{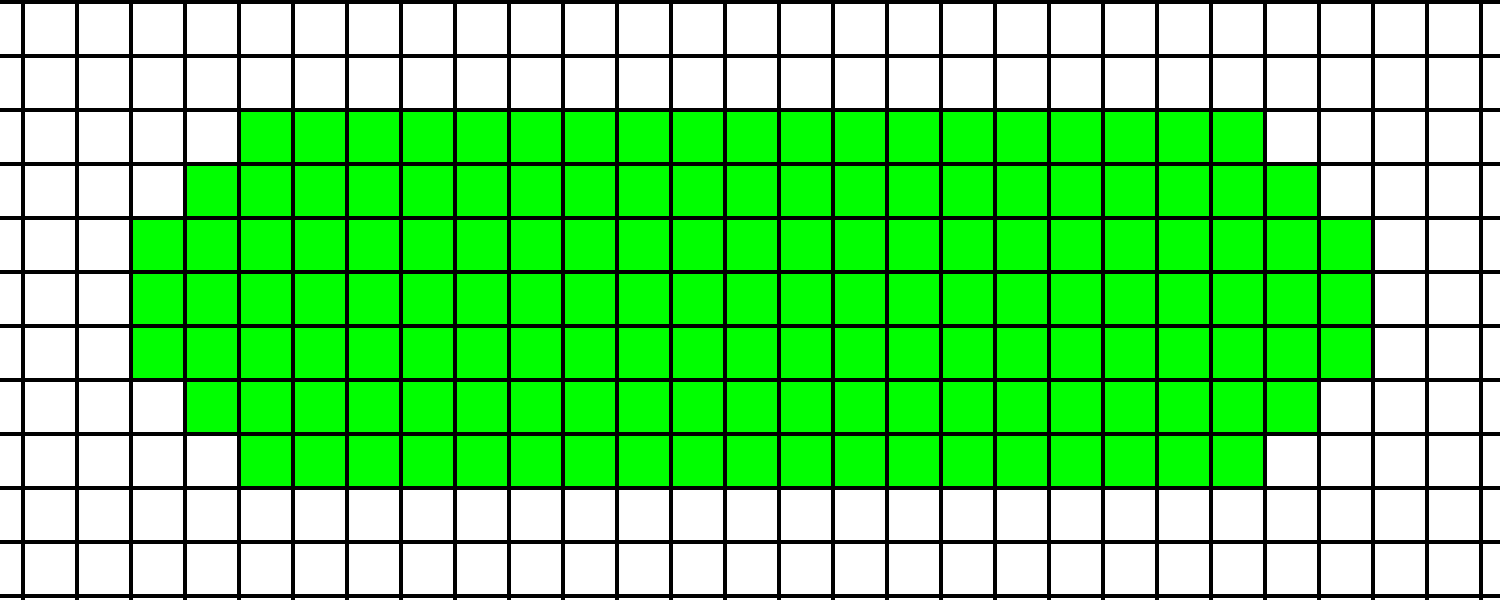}
             \vskip -5pt
             \caption*{(a)}
        \end{subfigure}
    \end{minipage}%
    \hskip 4pt
    \begin{minipage}{0.48\linewidth}
        \centering
        \begin{subfigure}
             \centering
             \includegraphics[width=1\linewidth]{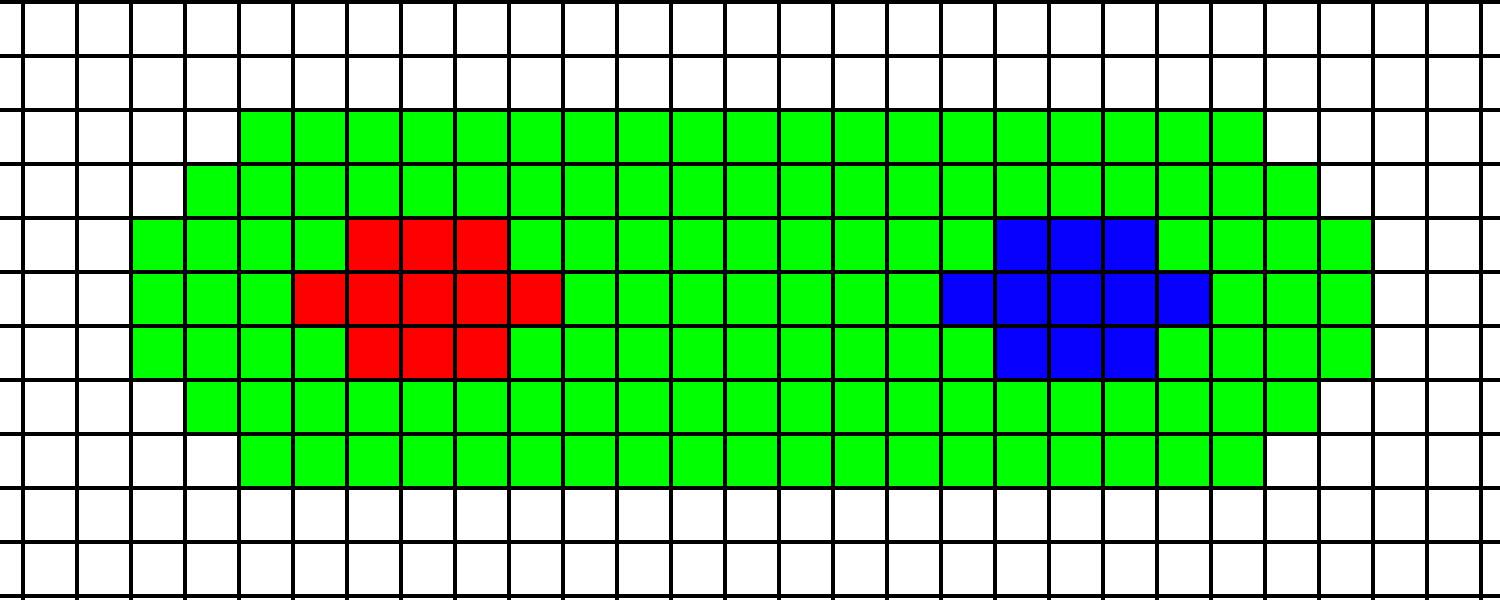}
             \vskip -5pt
             \caption*{(b)}
        \end{subfigure}
    \end{minipage}%
    \vskip 1pt
   \begin{minipage}{0.48\linewidth}
    \centering
        \begin{subfigure}
             \centering
             \includegraphics[width=1\linewidth]{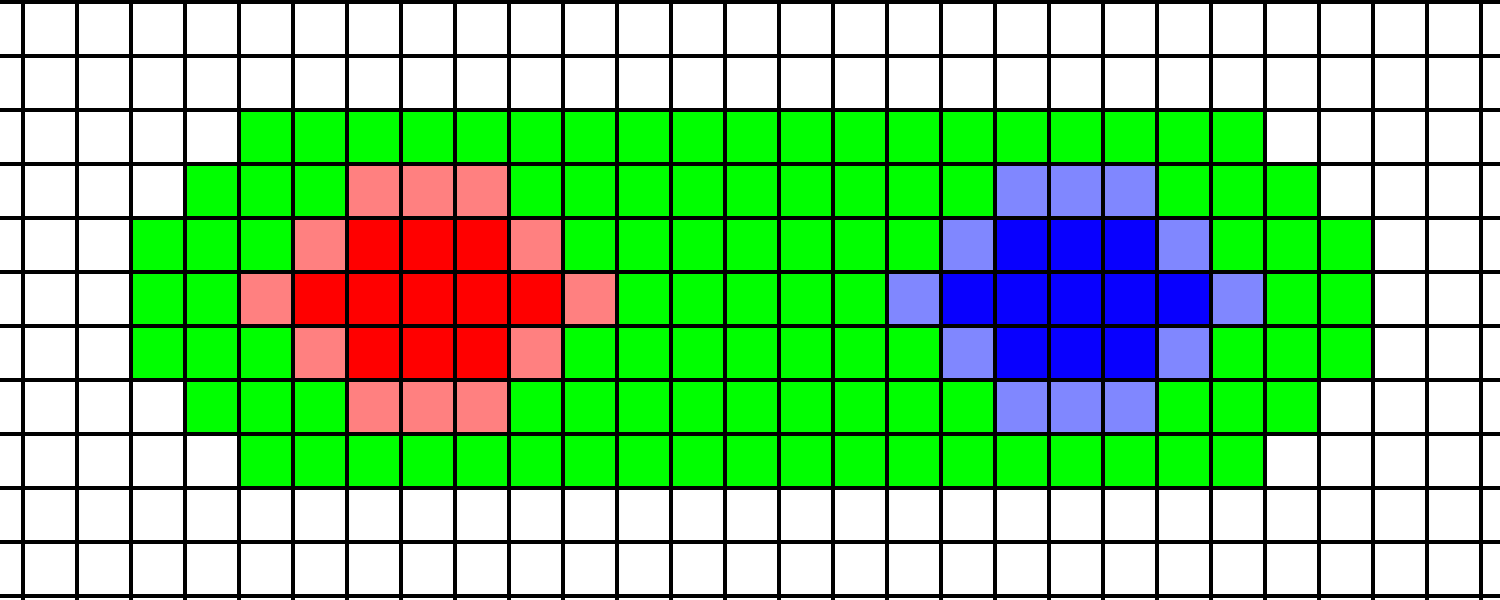}
             \vskip -5pt
             \caption*{(c)}
        \end{subfigure}
    \end{minipage}%
    \hskip 4pt
    \begin{minipage}{0.48\linewidth}
        \centering
        \begin{subfigure}
             \centering
             \includegraphics[width=1\linewidth]{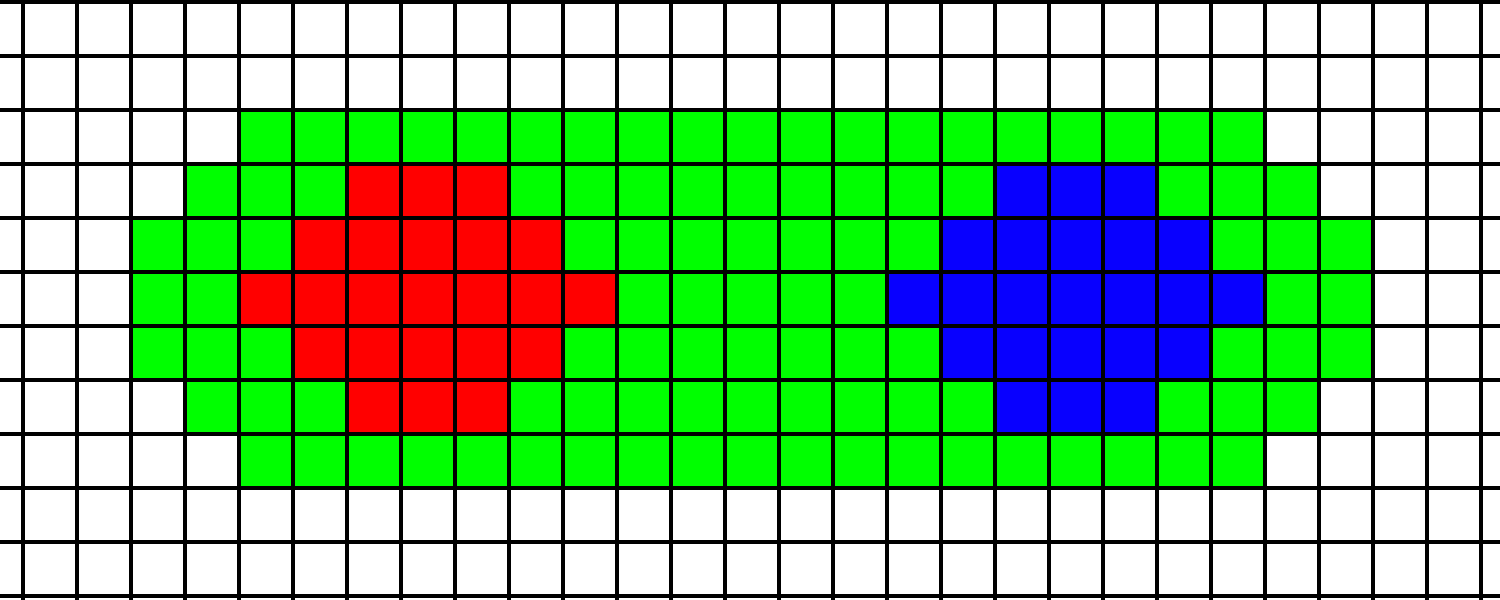}
             \vskip -5pt
             \caption*{(d)}
        \end{subfigure}
    \end{minipage}%
        \vskip 1pt
   \begin{minipage}{0.48\linewidth}
    \centering
        \begin{subfigure}
             \centering
             \includegraphics[width=1\linewidth]{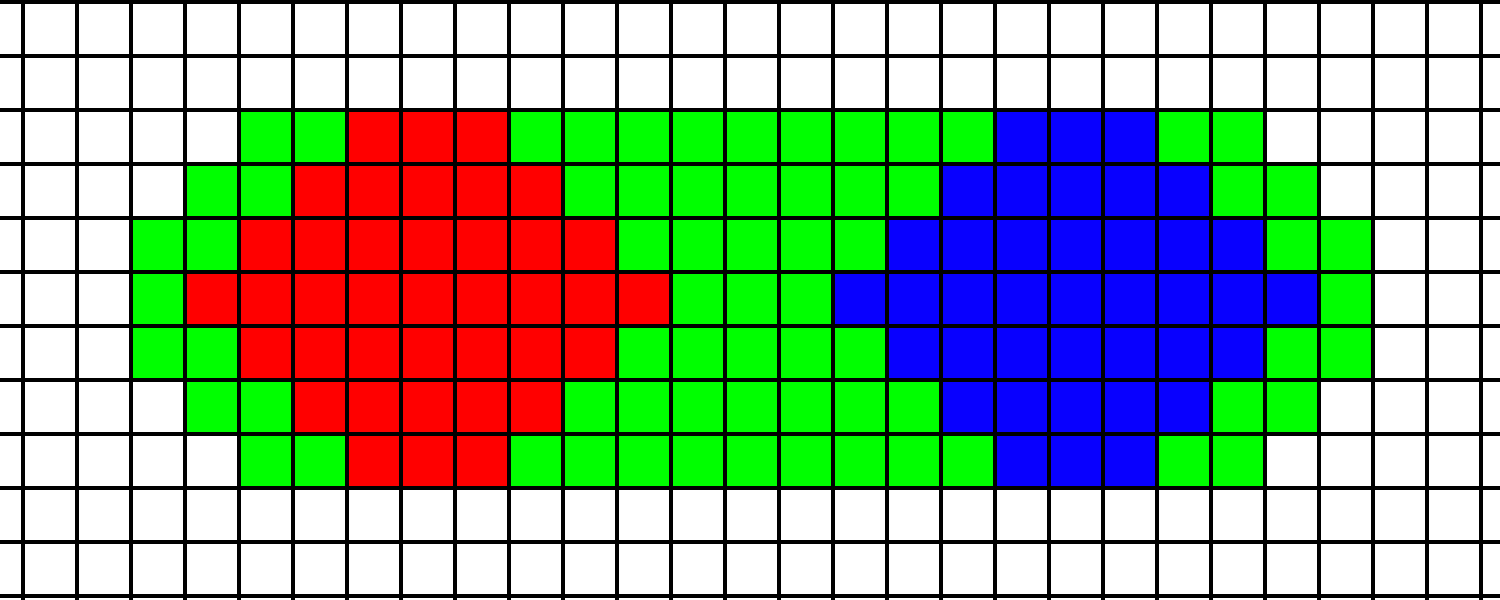}
             \vskip -5pt
             \caption*{(e)}
        \end{subfigure}
    \end{minipage}%
    \hskip 4pt
    \begin{minipage}{0.48\linewidth}
        \centering
        \begin{subfigure}
             \centering
             \includegraphics[width=1\linewidth]{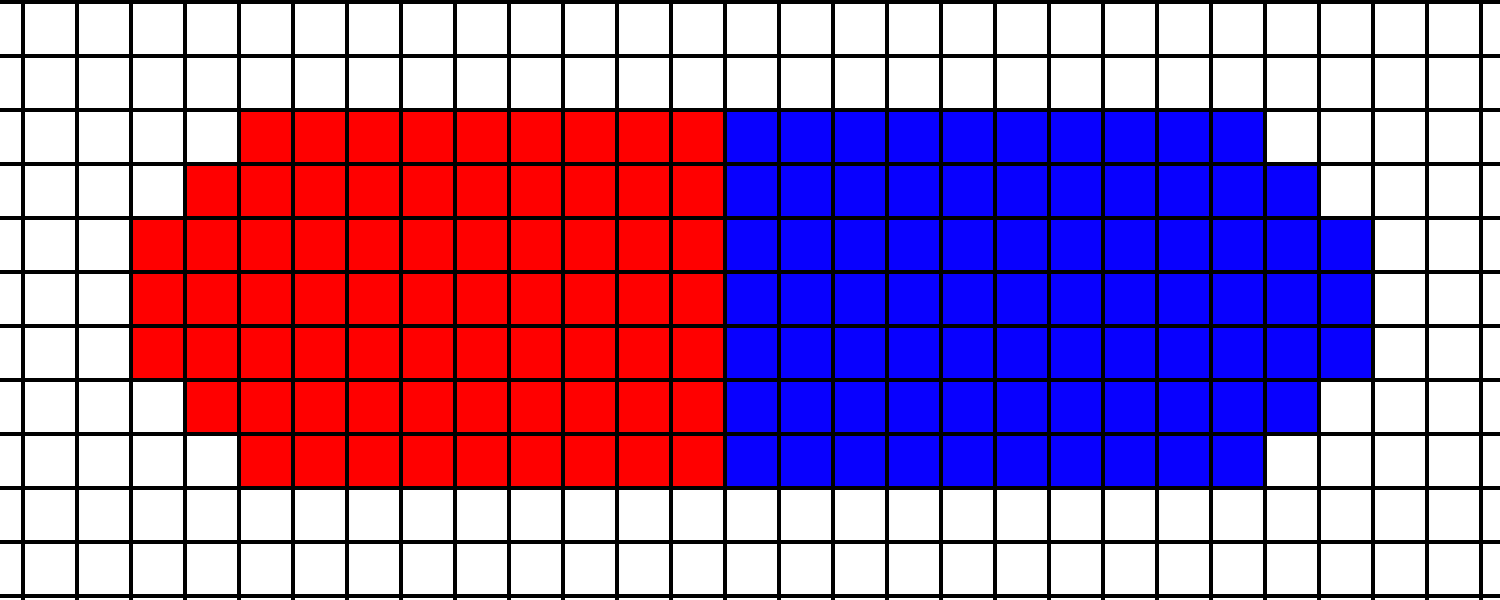}
             \vskip -5pt
             \caption*{(f)}
        \end{subfigure}
    \end{minipage}%
            \vskip 1pt
   \begin{minipage}{0.48\linewidth}
    \centering
        \begin{subfigure}
             \centering
             \includegraphics[width=1\linewidth]{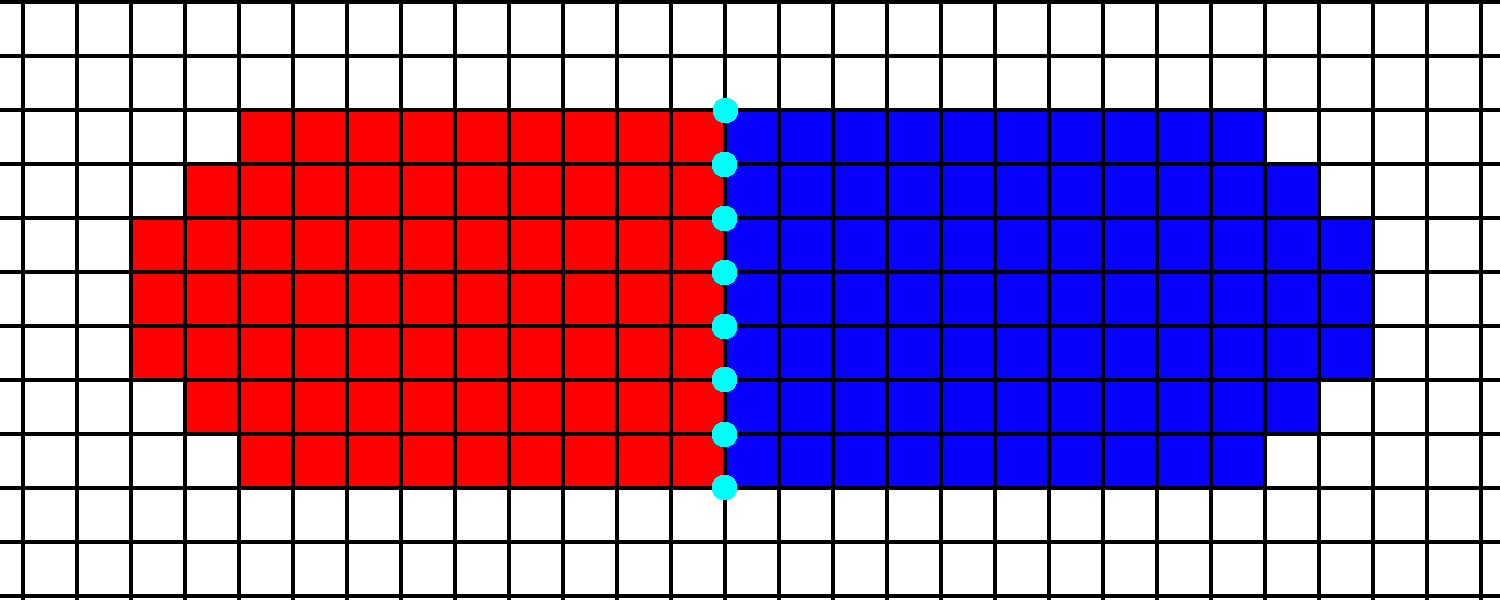}
             \vskip -5pt
             \caption*{(g)}
        \end{subfigure}
    \end{minipage}%
    \hskip 4pt
    \begin{minipage}{0.48\linewidth}
        \centering
        \begin{subfigure}
             \centering
             \includegraphics[width=1\linewidth]{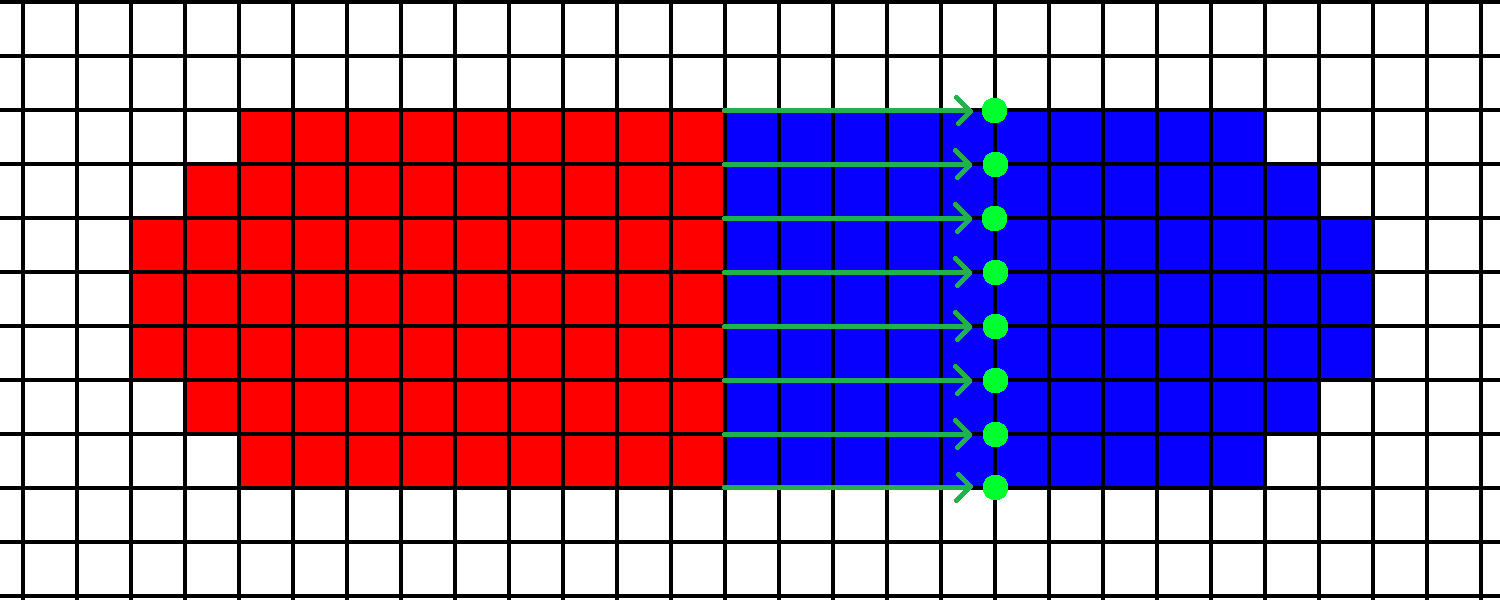}
             \vskip -5pt
             \caption*{(h)}
        \end{subfigure}
    \end{minipage}%
 \caption{This figure illustrates the splitting process of a vortical region using a 2D grid for ease of explanation. (a) shows a vortical region (green). (b) shows the initial contour tree-based segmentation in the form of the query segment (green) and seed segments (red and blue). (c) and (d) show one iteration of the layering. (c) shows the immediate neighboring cells (light red and light blue) of the seed segments in the query segment. (d) shows the end of the first iteration. (e) shows the end of the second iteration. (f) shows the final iteration of the layering where no cells are left in the query segment. (g) shows the boundary points (cyan). (h) shows the seed points (green) to compute vorticity lines which are five layers of cells away from the boundary points towards the bigger region.}
 \label{fig_1_1}
\end{figure}

\section{Additional Details}
In the following, we provide additional details for the methods discussed, such as Region Growing~\cite{adeelhairpin2023}, Splitting, and Layering. \cref{fig_1_1}(a) shows a vortical region. Using, e.g., $\lambda_2$, the vortical regions are extracted by finding the unstructured grids where the points fulfill the criterion that $\lambda_2 < 0$. However, in practice, $\lambda_2 < 0$ is less strict. In Region Growing~\cite{adeelhairpin2023}, we statistically find a threshold $\lambda_{2i}$ such that $\lambda_2 < \lambda_{2i}$, which is practically more stable. In \cref{fig_1_1}(b), the red and blue segments, and the green segment correspond to the two minima and maximum, respectively, of the minimal join tree-based segmentation explained in Section 3.1 of the main paper. The subsequent steps of the layering process are shown in \cref{fig_1_1}(c, d, e, f), respectively.

\begin{algorithm}[!t]
\caption{Vortex Separation Algorithm}
\label{alg:1}
\begin{algorithmic}
\small
\Procedure{Get\_Segments}{$vortexData$}
    \State \textbf{$\rightarrow$} Get critical point pairs with the highest persistence.
    \State $\rightarrow$ Perform topological simplification of $vortexData$.
    \State $\rightarrow$ $segmentIds\gets\ $Extract join tree and get segmentation.
    \State  \Comment{\textcolor{blue}{$segmentIds$ holds the segment ids for each cell}}
    \State \textbf{return} $segmentIds$
\EndProcedure
\Procedure{Layer}{$segmentIds,\ querySegment,\ seedSegments$}
    \State $\rightarrow$ $cellIds\gets$ neighbors of $seedSegments$ in $querySegment$.
    \State $\rightarrow$ Update $segmentIds$ of $cellIds$.
\EndProcedure
\Procedure{Fix\_Segments}{$vortexData,\ segmentIds$}
    \State $\rightarrow$ $querySegment\gets$ cells with the $segmentIds = -1$.
    \State $\rightarrow$ $seedSegments\gets$ cells with the $segmentIds \neq -1$.
    \State $\rightarrow$ $N\gets$ Number of cells in $querySegment$
    \While{$N\neq0$}
        \State $\rightarrow$ $\textsc{Layer}(segmentIds,\ querySegment,\ seedSegments)$
    \EndWhile
    \State \textbf{return} $segmentIds$
\EndProcedure
\Procedure{Avoid\_Split}{$vortexData,\ segmentIds$}
    \State $\rightarrow$ Find boundary points between the two segments.
    \State $\rightarrow$ Find seed points slightly away from the boundary points.
    \State $\rightarrow$ Extract streamlines using the $vorticity$ vector field.
    \State $\rightarrow$ $R_1=$ \# of cells overlapped by streamlines in segment 1.
    \If{$R_1 \geq 0.25$}
        \State \textbf{return} $true$
    \Else
        \State \textbf{return} $false$
    \EndIf
\EndProcedure
\Procedure{Separate}{$vortexData, regionIds$}
    \State $N\gets$ 0 \Comment{\textcolor{blue}{$N$ = number of regions}}
    \While{$N < 2$}
        \State $\rightarrow$ $segmentIds\gets\textsc{Get\_Segments}(vortexData)$
        \State $\rightarrow$ $segmentIds\gets\textsc{Fix\_Segments}(vortexData,\ segmentIds)$
        \State $\rightarrow$ $N\gets$ Number of unique ids in $segmentIds$
        \State $\rightarrow$ $avoidSplit \gets \textsc{Avoid\_Split}(vortexData,\ segmentIds)$
        \If{$avoidSplit$}
            \State \textbf{continue}
        \EndIf
    \EndWhile
    \State $\rightarrow$ Update $regionIds$ with $segmentIds$.
    \State $\rightarrow$ Get $newVortices[\ ]$ from $vortexData$ using $segmentIds$.
    \For{$newVortex$ in $newVortices$}
        \State $\rightarrow$ $\textsc{Separate}(newVortex,\ regionIds)$ \Comment{\textcolor{blue}{Recursive call}}
    \EndFor
\EndProcedure
\Procedure{Main}{$inputDataset$}
    \State $\rightarrow$ Use $Region\_growing()$ from~\cite{adeelhairpin2023} to get $vortexRegions$.
    \State $\rightarrow$ Initialize $regionIds$
    \State \Comment{\textcolor{blue}{$regionIds$ array keeps track of the vortex hierarchy}}
    \For{$disconnectedRegion$ in $vortexRegions$}
        \State $\rightarrow$ $\textsc{Separate}(disconnectedRegion,\ regionIds)$
        \State \Comment{\textcolor{blue}{Begin separation for each disconnected region}}
    \EndFor
\EndProcedure
\end{algorithmic}
\end{algorithm}

\begin{table*}[!t]
\caption{This table shows the performance statistics (\textit{in minutes(m)}) of the flow datasets used in our experiments. \emph{Segmentation Time} shows the performance of obtaining the initial contour tree-based segmentation. \emph{Layering Time} shows the time it takes to perform the layering. \emph{Avoid Split Time} is the time it takes to find the optimal seeds, extract vorticity lines, and decide whether to avoid the split. \emph{PP Time} is the time it takes for post-processing (PP) steps, including thresholding the new regions and recording the changes in the vortex hierarchy. \emph{Total Time} is the total time taken for the complete splitting process of the whole dataset. The performance is measured on a PC with an Intel Xeon(R) CPU E5-2640 without multiprocessing.}
\centering
\scriptsize
\begin{tabular}{|p{0.10\textwidth}|p{0.07\textwidth}|p{0.07\textwidth}|p{0.10\textwidth}|p{0.10\textwidth}|p{0.10\textwidth}|p{0.09\textwidth}|p{0.09\textwidth}| }
 \hline
 \multicolumn{8}{|c|}{Performance} \\
 \hline
  \multicolumn{1}{|c|}{Data Set} & \multicolumn{1}{c|}{\#Points} & \multicolumn{1}{c|}{\#Cells} & \multicolumn{1}{c|}{Segmentation Time (m)} & \multicolumn{1}{c|}{Layering Time (m)} & \multicolumn{1}{c|}{Avoid Split Time (m)} & \multicolumn{1}{c|}{PP Time (m)} & \multicolumn{1}{c|}{Total Time (m)}\\
 \hline
Couette Flow & 2169086 & 1580627 & 78 & 110 & 891 & 14 & 229\\
Channel Flow & 2476391 & 1570420 & 250 & 454 & 19 & 37 & 787\\
%Cylinder & -4.879 & 2.037 & 112822 & -0.004 & -0.1 & 12.12 & 119 & 66 & 13\\
%Crayfish & -0.086 & 0.026 & 100781 & -0.005 & -0.005 & 33.22 & 1087 & 397 & 9\\
%Plume & -51.544 & 13.69  &  560883 & -1 & -2 & 163.36 & 1281 & 442 & 14\\
%Couette & -0.728 & 0.488 &  28164288 & -0.01 & -0.05 & 1800 & 2651 & 1094 & 45\\
 \hline
\end{tabular}
\label{tab:Table1}
\end{table*}

\begin{figure}[!t]
 \centering 
    \begin{minipage}{0.90\linewidth}
    \centering
        \begin{subfigure}
             \centering
             \includegraphics[width=1\linewidth]{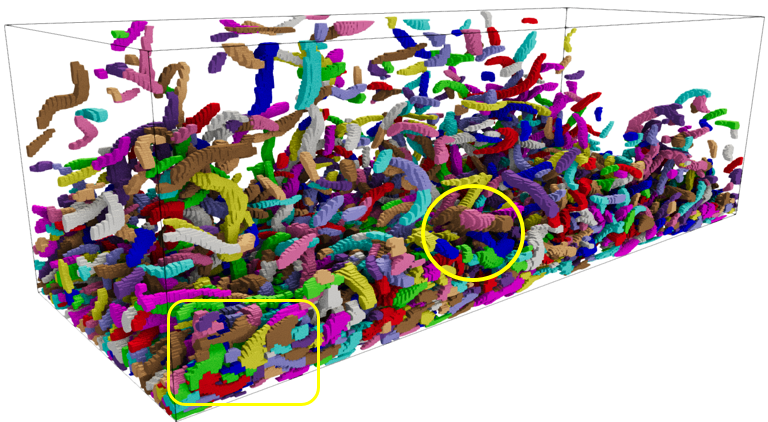}
             \vskip -20pt
             \caption*{(a)}
        \end{subfigure}
    \end{minipage}%
    \hskip 5pt
    \begin{minipage}{0.90\linewidth}
    \centering
        \begin{subfigure}
             \centering
             \includegraphics[width=1\linewidth]{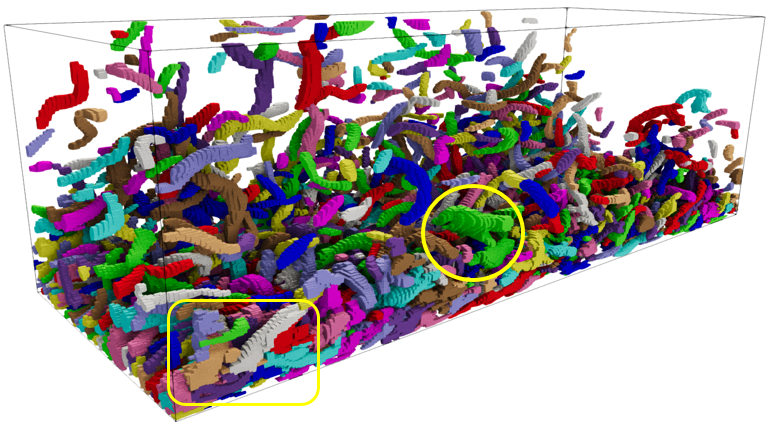}
             \vskip -20pt
             \caption*{(b)}
        \end{subfigure}
    \end{minipage}%
    \vskip 0.1pt
    \begin{minipage}{0.90\linewidth}
    \centering
        \begin{subfigure}
             \centering
             \includegraphics[width=1\linewidth]{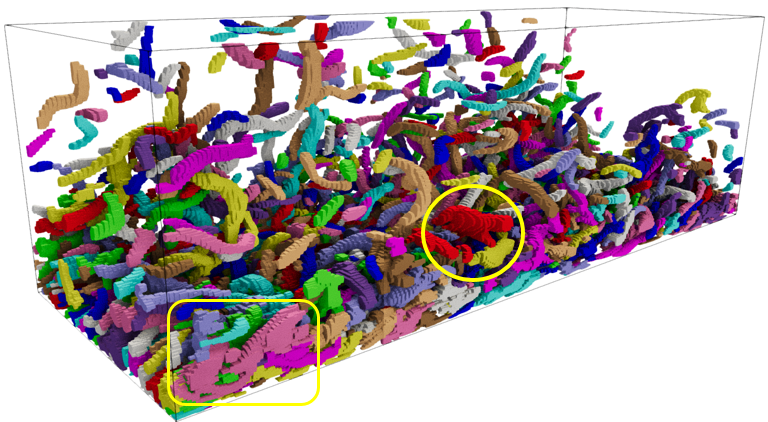}
             \vskip -20pt
             \caption*{(c)}
        \end{subfigure}
    \end{minipage}%
    \hskip 5pt
    \begin{minipage}{0.90\linewidth}
    \centering
        \begin{subfigure}
             \centering
             \includegraphics[width=1\linewidth]{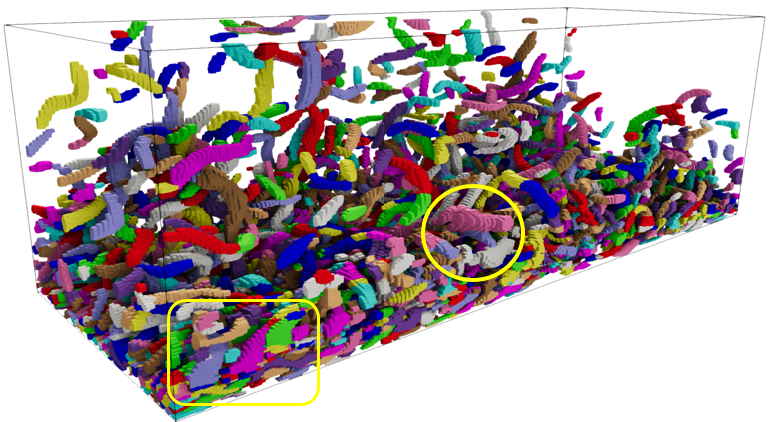}
             \vskip -20pt
             \caption*{(d)}
        \end{subfigure}
    \end{minipage}%
 \caption{(a) Zoomed in views of Fig. 9 (b, c, d, e) from the main text.}
 \label{fig_2_1}
\end{figure}

\section{Pseudo-Code for Vortex Separation}
The pseudo-code for the method is given in \cref{alg:1}. The algorithm starts with the \textsc{MAIN} function, where we first extract the vortical regions using the Region Growing approach~\cite{adeelhairpin2023}. \textit{regionIds} is an array that keeps track of the region id assigned to each cell. For each separate vortical region, we call the \textsc{SEPARATE} procedure. The parameter \textit{vortexData} represents the vortical region. In the \textsc{SEPARATE} procedure, we set \textsc{N}=0, where N is an integer which stores the number of new regions. The splitting process is done inside the \textbf{while} loop. We first get the initial contour tree-based segmentation using the \textsc{GET\_SEGMENTS} function. Then we call the \textsc{FIX\_SEGMENTS} function, which performs the layering process and splits the region into two new regions as illustrated in \cref{fig_1_1}. We store the number of new regions in the variable \textsc{N}. Then we call the \textsc{AVOID\_SPLIT} function that checks and decides if the region should be split or not. If 'Yes', the \textbf{while} loop ends and we record the changes in \textit{regionIds} and continue the recursive process on the new regions. If 'No', the next iteration of the loop starts, where the \textsc{GET\_SEGMENTS} function chooses the next (minimum-saddle) pair with the highest persistence. If we have exhausted all the (minimum-saddle) pairs, then the \textbf{while} loop ends, terminating the process for the current region.

\section{Additional Results}
\cref{fig_2_1} shows the zoomed-in views of the vortex separation results on the channel flow dataset from Fig. 9 of the main paper. \cref{tab:Table1} shows the performance of the algorithm in the current implementation of the method. We will improve this implementation to make it practical for large-scale datasets.

% \section*{Figure Credits}
% \Cref{fig:hairpin_stages} is taken from \cite{Jiji2009}.

\acknowledgments{We thank the anonymous reviewers for their valuable feedback. This research was supported by NSF OAC 2102761.}

\bibliographystyle{abbrv-doi-hyperref}
%\bibliographystyle{abbrv-doi-hyperref-narrow}
%\bibliographystyle{abbrv-doi}
%\bibliographystyle{abbrv-doi-narrow}

% \bibliography{template}
\bibliography{main}

%% ^^^^^   FOR IEEE VIS, EVERYTHING HERE MAY BE INCLUDED IN THE    ^^^^^ %%
%% 2-PAGE ALLOTMENT FOR REFERENCES, FIGURE CREDITS, AND ACKNOWLEDGEMENTS %%